\documentclass[12pt]{iopart} 
\usepackage{amssymb,latexsym} \usepackage{txfonts} 
\usepackage{xcolor}
\usepackage[pdftex]{graphicx} \usepackage[square, sort&compress,
  numbers]{natbib} \usepackage[toc,page]{appendix} \usepackage{bm}
\definecolor{darkgreen}{HTML}{008000}
\definecolor{magenta}{HTML}{CC00CC}
\definecolor{cyan}{HTML}{00FFFF}
\definecolor{darkbrown}{rgb}{0.7,0.3,0}

\begin{document}


  
  \title{Bubble propagation in  Hele-Shaw channels with centred constrictions}

\author{Andr\'es Franco-G\'omez\textit{$^{1}$}, Alice
  B. Thompson\textit{$^{2}$},  Andrew L. Hazel\textit{$^{2}$} and
  Anne Juel\textit{$^{1}$}\footnote{Corresponding author:
    anne.juel@manchester.ac.uk}}

\address{$^1$ Manchester Centre for Nonlinear Dynamics \& School of
  Physics \& Astronomy, The University of Manchester, Manchester M13
  9PL, UK.}  \address{$^2$ Manchester Centre for Nonlinear Dynamics \&
  School of Mathematics, The University of Manchester, Manchester M13
  9PL, UK.}
\vspace{10pt}
\begin{indented}
  \item[] \today
\end{indented}

\begin{abstract}
We study the propagation of finite bubbles in a Hele-Shaw channel,
where a centred occlusion (termed a rail) is introduced to provide a small axially-uniform
depth constriction. For bubbles wide enough to span the channel, the
system's behaviour is similar to that of semi-infinite fingers and a symmetric
static solution is stable. Here, we focus on smaller bubbles, in
which case the symmetric static solution is unstable and the static
bubble is displaced towards one of the deeper regions of the channel on
either side of the rail. Using a combination
of experiments and numerical simulations of a depth-averaged model, we
show that  a bubble propagating axially due to a small imposed flow rate can be stabilised in a
steady symmetric mode centred on the rail through a
subtle interaction between stabilising viscous forces and
destabilising surface tension forces. However,  for sufficiently large
capillary numbers $Ca$, the ratio of viscous to surface tension
forces, viscous forces in turn become destabilising thus returning the
bubble to an off-centred propagation regime. With decreasing bubble
size, the range of $Ca$ for which steady centred propagation is stable
decreases, and eventually vanishes through the coalescence of two
supercritical pitchfork bifurcations. The depth-averaged model is
found to accurately predict all the steady modes of propagation
observed experimentally, and provides a comprehensive picture of the
underlying steady bifurcation structure. However, for sufficiently
large imposed flow rates, we find that initially centred bubbles do
not converge onto a steady mode of propagation. Instead they
transiently explore weakly unstable steady modes, an evolution which
results in their break-up and eventual settling into a steady
propagating state of changed topology. 
\end{abstract}

%
%
%
%
%

\section{Introduction}


Understanding the motion of gas bubbles 
within liquid-filled vessels is a fundamental problem in fluid
mechanics, and the interaction between the 
bounding geometry and deformable gas-bubble can provoke non-trivial
dynamics that is quite different from that of bubbles in 
unbounded flow \cite{Baroud2010}. 
Pure hydrodynamic interactions induce 
migration of the bubbles normal to the predominant direction of flow and they
accumulate at particular locations within the vessel's
cross-section \cite{Stan_etal2011}. Although our principal motivation
for this study is 
fundamental fluid-mechanical interest, a detailed understanding of these
effects can be used to develop passive
bubble sorting applications in lab-on-a-chip devices 
\cite{Sajeesh2014}.
Migratory mechanisms also operate on particles and droplets and 
the resulting applications in microfluidics 
include assays for chemical and biological dynamics
\cite{Dressler2017}, segregation of blood components
\cite{DiCarlo_etal2007}, trapping of micro-organisms in
water \cite{Kim_etal2014}, and formation and 
purification of emulsions or colloids \cite{Hashimoto2007}.

 Passive bubble sorting requires bubbles of different properties to be
transported at different positions within the vessel's cross-section. It is
then straightforward to use geometric separators to collect bubbles
with the required properties.
The accuracy and sensitivity of these
methods will both be maximised if
the cross-sectional position of a bubble can be significantly 
altered in response to small changes in governing parameters:
a situation that arises naturally 
if the positional migration is a consequence of changes in the stability
and number of bubble propagation modes, \textsl{i.\@e.\@} near
bifurcation points of the underlying dynamical system. 
Thus, nonlinearity of the system is essential for such sorting devices.
For sufficiently small vessels, inertial effects in the
liquid are often negligible and the
nonlinear effects arise only through the possible deformation of the gas-liquid interface.

If there are no inertial effects in the fluid then rigid particles do 
not migrate within the cross section \cite{CoxMasonReview1971},
whereas the deformability of droplets and bubbles allows them to migrate
\cite{LealReview1980,Stan_etal2011}
through a subtle interplay between surface tension and viscous forces, whose
relative importance can be quantified by a capillary number, $Ca = \mu^*
U_b^*/\sigma^*$, where $\mu^*$ is the viscosity of the suspending
fluid, $U^*_b$ is the bubble velocity and $\sigma^*$ is the surface
tension at the interface between the suspending liquid and gas
bubble. For pressure-driven flow in vessels with circular,
annular or rectangular cross-sections, a deformable object 
will always migrate towards the centre of the cross-section
\cite{KarnisMason1967,TanveerSaffman1987}. Thus, the system must be
modified in some fashion in order to induce multiple propagation
modes and hence to allow robust passive sorting. 

 In the absence of other external forcing, the only possible 
modifications are changes to the vessel geometry. For example, Abbyad \textsl{et
al.} \cite{Abbyad2011} showed that drops may be guided to
desired locations in a microchannel by inscribing grooves in its top
surface. These grooves anchor the droplets by enabling them to expand to
reduce their surface energy and the grooves thus passively guide the droplets
along pre-determined paths for moderate flow rates. 
Furthermore, multiple grooves of different widths have
been used to sort droplets by their size or capillary number
\cite{Yoon_etal2014}.
Although such grooves must be specifically manufactured and hence are typically used as passive control mechanisms,
droplets can also be actively moved between
grooves by use of cheap commercially available DVD lasers \cite{Frot2016}. 

In this paper, we consider the propagation of finite gas bubbles suspended
within silicone oil driven through a vessel of large-aspect-ratio rectangular
cross-section, a Hele-Shaw channel,
that we modify by the addition of a centred localised height
constriction, henceforth termed a ``rail''. 
Two-phase flow in Hele-Shaw channels in the absence of  such rails 
has been extensively studied
since the seminal papers of Saffman \& Taylor
\cite{SaffmanTaylor1958,TaylorSaffman1959}. 
In these systems, depth-averaged models predict a variety of possible
bubble propagation modes \cite{Tanveer1987,Greenetal2017}, but for
non-zero surface tension only
one, centred, propagation mode is stable \cite{TanveerSaffman1987}. 

Previous studies of two-phase displacement flows in channels constricted by
a rail have focused on the propagation of open bubbles, often
referred to as fingers, in channels of both small
\cite{DeLozar2009,Pailha2012,Hazel2013} and large
\cite{Thompson2014,FrancoGomez2016} cross-sectional 
aspect ratios, and closed bubbles longer than the width of the channel \cite{Jisiou2014}. These studies have
uncovered a wide variety of steady and oscillatory propagation modes:
at low flow rates surface tension dominates and a finger will 
propagate centrally along the channel; at high flow rates the fingers
tend to propagate asymmetrically along one side of the rail, which avoids
displacing fluid within the constricted region in which the local viscous
resistance is greatest. For intermediate flow rates a variety of
centred and asymmetric oscillatory solutions have been observed.

In an unoccluded channel, bubbles with finite volume have a number of features in common with air fingers; multiple solution branches exist, but only a single, centred branch is stable. Franco-G\'omez et al. 
\cite{FrancoGomez2017} recently studied the behaviour of small bubbles in the presence of a rail with width comparable to the bubble diameter and found that these bubbles can exhibit
bistability of centred and off-centred propagation modes inside a tongue-shaped region of
the parameter space spanning bubble size and imposed flow rate; this bistability
can in principle be used as a mechanism for passive sorting of bubbles by size.
These small bubbles remain nearly circular even during migration across the channel.

For the present work, we consider the transition in behaviour as bubble size varies from diameters comparable to the rail width to diameters comparable to the channel width, which bridges the two recent studies of Franco-G\'omez et al. \cite{FrancoGomez2016,FrancoGomez2017}.
We shall concentrate on the fundamental questions of 
how the number and stability of different
propagation modes vary with changes in flow rate and bubble
volume for a given rail geometry.
We find that viscous forces can act to either stabilise or
destablise bubbles propagating in a centred configuration over the
rail depending on the value of $Ca$ and the size of the bubble. 
For large flow rates, we find that an initially centred 
bubble selects neither a 
steady symmetric nor steady asymmetric mode of propagation, but exhibits complex
unsteady behaviour, with significant shape deformation, that appears to be organised by unstable steadily
propagating solutions of the system. Unstable invariant solutions
have recently been recognised as playing a key role in the transition
to turbulence in shear flows \cite{Kawahara2012,barkley_2016} and we
conjecture that they may play a similar role in the transition to
the disordered behaviour that we have observed in this system.

The paper is structured as follows. The experimental and numerical
methods are presented in sections \ref{expmeth} and \ref{nummeth},
respectively. The steady modes of propagation of a finite bubble are
discussed in section \ref{sec1} through the comparison between
experiment and numerical simulation. We find that a centred
mode of propagation disappears with decreasing bubble size, which
 is presented in
section \ref{sec2}. The bifurcation structure underlying steady bubble
propagation is established through numerical simulations in section \ref{sec3}
and used to interpret the unsteady propagation of bubbles at large
flow rates in section \ref{sec4}. Conclusions are given in section
\ref{conc}.

\section{Experimental methods}\label{expmeth}

  \begin{figure}[h]
    \begin{center}
      \includegraphics[width=\textwidth]{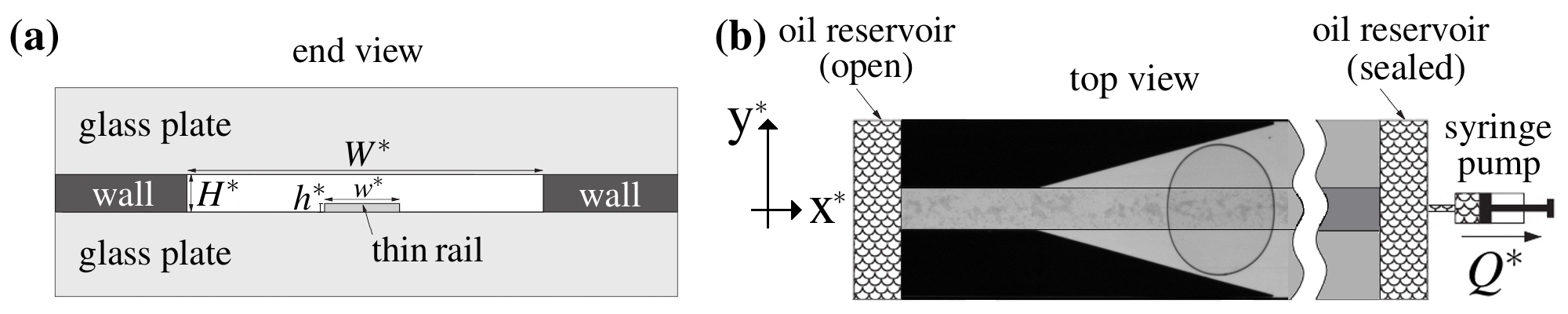}
    \end{center}
    \vspace{-0.5cm}
    \caption{\small  Schematic diagram of the experimental system. (a)
      Cross-sectional view of the channel (width $W^*=30.0$ mm and
      height $H^*=1.0$~mm) with a prescribed depth variation: an
      axially uniform, thin rail (width $w^*=7.0$ mm and height
      $h^*=24.0$ $\mu$m) is positioned on the bottom boundary,
      symmetrically about the centreline of the channel. (b) Top view
      of the experimental system, showing the constricted channel
      upstream of the flow channel used to position the bubble in a
      centred static initial
      configuration.}\label{fig:Setup_BubbleIni}
  \end{figure}

A schematic diagram of the experimental setup is shown in Figure
\ref{fig:Setup_BubbleIni}. It is described in detail in previous work
\cite{FrancoGomez2016,FrancoGomez2017}, and thus, we only provide a
brief description of the points pertinent to the present study. All
dimensional quantities are starred. The experimental channel consisted
of two parallel float glass plates of dimensions $60$ cm $\times$ $10$
cm $\times$ $2$ cm separated by brass sheets of uniform height
$H^*=1.0$~mm, accurate to within $0.1\%$ (Figure
\ref{fig:Setup_BubbleIni}a). The width of the channel was $W^*=30.0
\pm 0.1$~mm, so that its aspect ratio was $\alpha=W^*/H^*=30$ (with
the exception of section \ref{sec4}, where we also consider a channel
with $W^*=40$~mm so that $\alpha=40$). A thin rail of  uniform width
$w^*=7.0\pm0.1$~mm (and also $w^*=10.0 \pm 0.1$~mm in section
\ref{sec4}), and thickness  $h^*=24\pm1$ $\mu$m ($2.4$ \% of the
height of the cell) was cut from a sheet of polypropylene film, and
bonded to the bottom plate of the channel with its centreline aligned
along the centreline of the channel. Measurements of the rail profile
using a step profilometer (DekTak II) are presented in the supplementary material of \cite{FrancoGomez2017}. The
widths of the two regions between the edge of the rail and the side
walls of the channel were carefully calibrated to differ by less than
$1$ \% ($<100$ $\mu$m), and the channel was levelled horizontally to
better than $0.05$ $^{\circ}$.

Perspex fluid reservoirs with a small bleed hole on their upper
surface were attached to both ends of the channel, with a tight seal
achieved by using rubber gaskets (Figure
\ref{fig:Setup_BubbleIni}b). The channel was completely filled with
silicone oil (Basildon Chemicals Ltd), which had viscosity
$\mu^*=1.0\times10^{-2}$ Pa s, density $\rho^*=961$ kg m$^{-3}$ and
surface tension $\sigma^*=2.1\times10^{-2}$ N m$^{-1}$ at the
laboratory temperature of $21$ $^\circ$C.  One of the reservoirs was
filled to a level higher than the upper boundary of the channel end
and left open to the atmosphere. The other reservoir was completely
filled, sealed and was connected to a syringe pump (Legato200 series),
which was used to impose steady flow in the channel by withdrawing oil
at a constant flow rate $Q^*$.  In order to form a finite air bubble
inside the channel, the level of oil in the open reservoir was reduced
so that  air could be infused into the channel upon withdrawal  at low
flow rate of a small volume of oil from the other end of the
channel. The open reservoir was then refilled through its bleed hole
to avoid further air entering the system. Moreover, the size of the
bubble could be decreased by positioning the edge of the bubble at the
upstream end of the channel and allowing a small volume of air to
escape through the bleed hole in the reservoir upon infusion of liquid
at low flow rate. Air fingers were generated by keeping a low volume
of oil in the open reservoir and fully opening the bleed hole so that air
continuously entered the channel on removal of oil via the syringe pump.

Prior to propagation at constant flow rate $Q^*$, the bubble was
placed in a 5~cm long constricted channel that spanned the width of
the rail at the upstream end of the channel (Figure
\ref{fig:Setup_BubbleIni}b). This channel expanded linearly over a
length of 4~cm to reach the width of the flow channel. The bubble was
propagated with a low flow rate $Q^*=2.0$ ml/min into the diverging
channel in order to allow the shape of the bubble to relax before
switching off the flow for a few seconds so that the initial position
of the bubble was centred and static. Note that the rail is placed
along the entire length of the channel, including the transition region. All experiments were performed
using this initial condition following preliminary tests showing that
similar states of propagation were obtained for moderate flow rates
regardless of the initial configuration. The mode of bubble propagation was found to depend on initial conditions only for the high flow rates explored in section \ref{sec4}. After each
experiment, the bubble was propagated back to the inlet channel, by
infusing liquid into the channel using the syringe pump. The
experimental protocol was automated to enable the steady modes of
bubble propagation to be characterised over a range of flow rates of
$1\leq Q^*\leq 70$ ml/min by incrementing the flow rate in steps of
$1$~ml/min . During these experiments, the bubble volume remained
constant to within $\pm 1 \%$ \cite{FrancoGomez2017}.

Bubble propagation was monitored using a Dalsa Genie TS-M3500 camera
with a $35$ mm $f$/1.4 lens (Carl Zeiss T$^*$ Distagon) mounted above
the horizontal plane of the channel at a distance of $0.94$ m, which
captured images of $1920$ $\times$ $218$ pixels corresponding to
$271.7$ mm $\times$ $30.8$ mm (resolution of $141.5$ $\mu$m/px). The
experiment was back-lit with a custom made LED light box made of
diffusive perspex (opal 070), which produced uniform white light and
was placed under the channel. Refraction of light at the bubble
interface made the contour of the projected area of the bubble in top
view appear dark in a light background. Sequences of $140\leq
N\leq600$ images were captured at frames rates between $10\geq f^*\geq
1$ frames per second, respectively, for flow rates $1\leq Q^*\leq 270$
ml/min. Both the syringe pump and the camera were
controlled in LabView; the images were
processed using MATLAB R2014a routines.  The
edge of the bubble contour was extracted from the grey-scale
images, and these binary bubble profiles were used to measure the area
of the bubble in top view $A^*$. The size of the bubble was quantified
by its effective static diameter $D^*$, defined as $D^*=2
\sqrt{A_0^*/\pi}$, which was obtained from its area $A_0^*$,  measured
when the bubble was in its static initial position. Hereinafter, we
will refer to the static effective diameter relative to the width of
the channel, $D=D^*/W^*$. The range of bubble sizes investigated in this paper is
$0.35 \le D \le 0.87$. The smallest bubbles considered here lie towards the
upper end of the range considered by Franco-G\'omez
\textsl{et al.}
\cite{FrancoGomez2016}, where $D$ was in the range $0.18 \leq D\leq 0.49$. Note however that the results in \cite{FrancoGomez2016} were characterised by the ratio of bubble diameter to rail width: $d = D^{*}/w^{*}$. In terms of $d$, the present range of bubble sizes is $1.5 \le d \le 3.73$, while \cite{FrancoGomez2016} explored the range $0.77 \le d \le 1.81$.

The position of the propagating bubble was
quantified by the displacement of the centroid of its top-view contour
relative to the centreline of the channel, $y_c^*$, which takes
non-dimensional values $\displaystyle -1 < y_c=2y^*_c/W^* <1$, and by the axial position of its tip: the point on the interface with the largest $x$-coordinate.  The
instantaneous speed of the bubble tip, $U_b^*$, was determined by
dividing the axial distance covered by the bubble tip between frames by the elapsed time, which was monitored
for each frame.  For most experiments in this paper, bubble deformation was mild, and states were classified as steady if both $U_b^*$ and $y_c^*$ had reached constant values by the end of the visualisation window; otherwise bubble propagation was classified as unsteady.

Steady modes of propagation will be characterised in terms of the
capillary number $Ca = \mu^* U_b^* / \sigma^*$. However, for unsteady modes of propagation, the bubble speed may vary in time, and so the unsteady results discussed in section \ref{sec4} will be presented in terms
of the imposed  non-dimensional flow rate defined as $Q=\mu^* U^*_{\rm 0(exp)} /
\sigma^*$,  where $U^{*}_{\rm 0(exp)}=Q^* / (W^*H^*)$ is the average
speed of the fluid flow within the channel in the absence of the
rail. The ratio between gravitational and surface tension forces was
$Bo=\rho^* g^*H^{*2}/(4\sigma^*)=0.11$, where $g^*$ is the acceleration
due to gravity. This low value indicates that buoyancy did not affect
bubble propagation significantly \cite{FrancoGomez2016}. 
The importance of inertial terms relative to viscous forces is governed by the reduced Reynolds number $Re = \rho^* U_{\rm 0(exp)}^* H^{*2}/(\mu^* W^*)$, which has maximum value $0.2$; thus inertia is negligible.

\section{Depth-averaged model for bubble propagation}\label{nummeth}

 \begin{figure}[ht]
    \begin{center}
      \includegraphics[width=12.5cm]{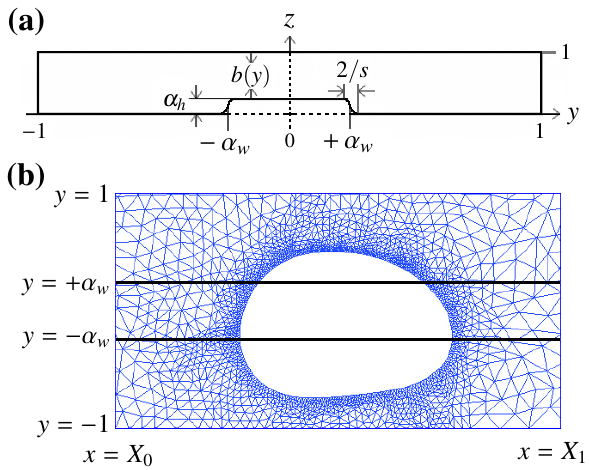}
    \end{center}
    \vspace{-0.5cm}
    \caption{\small (a) Cross-section of the channel with a prescribed
      depth profile $b(y)$ where $\alpha_w=w^*/W^*=0.288$ is the
      fractional rail width, $\alpha_h=h^*/H^*=0.024$ the fractional
      rail height and $2/s$ the width of the profile edge, with
      $s=40$. (b) Triangular mesh and bubble contour implemented in
      the numerical model. Solid black lines at $+\alpha_w$ and
      $-\alpha_w$ denote the edges of the rail. Figures (a) and (b)
      are plotted in a dimensionless coordinate
      system.}\label{fig:BubbleMeshRailProf}
  \end{figure}
  
We model the motion of finite bubbles in a constricted channel using
the same depth-averaged model
as Franco-G\'omez \textsl{et al.} \cite{FrancoGomez2017}, itself adapted from
our previous models developed
for air fingers \cite{Thompson2014,FrancoGomez2016} and found to be
quantitatively accurate for $\alpha = W^*/H^* \geq 40$.

The geometry of the channel cross-section used in the model is
shown in Figure \ref{fig:BubbleMeshRailProf}a. We introduce a
Cartesian coordinate system aligned with the channel such that the
in-plane coordinates $x$ and $y$ have been non-dimensionalised by
$W^*/2$, whereas the $z$ coordinate spanning the height of the
cross-section has been non-dimensionalised by $H^*$. The rail is
represented by a smooth tanh-profile with height given by, 

  \begin{equation}\label{DepthProfile}
     b(y) = 1 - \frac{\alpha_h}{2}[\tanh s(y+\alpha_w)-\tanh
       s(y-\alpha_w)],
  \end{equation}
where $s$ is the sharpness of profile edges, and $\alpha_h=h^*/H^*$,
$\alpha_w=w^*/W^*$ are the fractional height and fractional width
of the profile, respectively.

 The rail parameters used in the model are the same as those chosen based on profilometry measurements  
 by Franco-G\'omez \textsl{et
 al.} \cite{FrancoGomez2017}  and used for their study of small bubble
 propagation: $\alpha_h=0.024$ and
 $\alpha_w=0.288$.
 Following \cite{FrancoGomez2016,FrancoGomez2017},
 we performed the simulations
 with a sharpness parameter $s=40$.
Here $\alpha_h=0.024$ is equal to the measured value of $h^*/H^*$, however $\alpha_w$ is larger than $w^*/W^*$. As discussed in \ref{A2}, if $s=40$, the choice $\alpha_w=0.288$ ensures that the width of the top surface of the smoothed rail matches the width of the top surface of the experimental rail.
We provide a sensitivity
 analysis of the effects of rail width and sharpness in \ref{A2}. 

The action of the syringe pump is modelled by imposing a constant
pressure gradient $-G^*\bf{e}_x$ far ahead of the bubble tip, where $G^*$ takes the value necessary so that the dimensional volume flux is fixed at $Q^*$.
We non-dimensionalise the two components of the depth-averaged horizontal velocity $\mathbf{u}^*$ on the scale $U_0^* = Q^*/(H^* W^*)$, horizontal coordinates on the scale $W^*/2$, time on the scale $W^*/(2 Q^*)$ and pressure on the scale $6 \mu Q^*/H^{*3}$, where $\mu^*$ is the dynamic viscosity of the fluid.
%
%
After
applying the lubrication approximation \cite{Reynolds1886}, the
governing equation for the viscous, incompressible fluid in the frame
of reference moving with speed $\mathbf{U}_b = (U_b,0)$,
where $U_{b} = U_{b}^{*}/U_{0}^{*}$, is
 
 \begin{equation}\label{LaplaceEq}
     \nabla\cdot(b(y)^3\nabla p)=0 \quad \mathrm{in}\quad  \Omega,
 \end{equation}
  where $\Omega$ denotes the fluid domain. The fluid domain is given
  by $X_0
  \leq x \leq X_1$, $-1 \leq  y \leq 1$, excluding the region occupied
  by the bubble,
  where $X_0$ and $X_1$ are truncation coordinates behind and
  ahead of the centre of the bubble (Figure \ref{fig:BubbleMeshRailProf}b). 

The conditions at the bubble interface ${\bf R}=(x,y)$ and on the
channel boundaries are
  \begin{equation}\label{VeloCont}
       {\bf \hat  n}\cdot \frac{\partial \mathbf{R}}{\partial t}+{\bf
         \hat n}\cdot \mathbf{U}_b+b^2{\bf \hat n}\cdot \bm{\nabla}
       p=0\quad \mathrm{on}\quad  \partial \Omega_b,
  \end{equation}
  \begin{equation}\label{PressureJump}
       p_{b}-p=\frac{U_b}{3\alpha
         Ca}\left(\frac{1}{b(y)}+\frac{\kappa}{\alpha}\right)\quad
       \mathrm{on}\quad  \partial \Omega_b,
  \end{equation}
  \begin{equation}
  \label{NoNormalVelocity}
 \frac{\partial p}{\partial y}=0\quad \mathrm{on}\quad y=\pm 1, 
  \end{equation} 
    \begin{equation}
  \frac{\partial p}{\partial x}= -G\quad \mathrm{on}\quad x=X_0,   \quad\quad \frac{G}{2} \int_{-1}^1  b^3(y)\, \mathrm{d}y = 1. 
  \end{equation} 

  \begin{equation}\label{PressWallEnds}      
      p=0\quad \mathrm{on}\quad x=X_1,
  \end{equation}
  where $\partial \Omega_b$ denotes the bubble boundary with unit
  normal $\bf \hat n$ directed into the fluid. The interface
  conditions in the moving frame have been previously derived
  \cite{FrancoGomez2016} and the only difference from the interface
  conditions in the fixed lab frame \cite{Thompson2014}, is in
  equation (\ref{VeloCont}), which adds the velocity of the frame, ${\bf u}=-\mathbf{U}_b-b^2
  \bm{\nabla} p$, into the kinematic boundary condition $\mathbf{\hat{n}}\cdot \partial {\bf R}/\partial t = \mathbf{\hat{n}} \cdot \mathbf{u}$. Note that (\ref{NoNormalVelocity}) implies that there is no normal velocity at the channel side walls.

The $\partial {\bf R}/\partial t$ term in (\ref{VeloCont}) is the only time derivative in
the problem and prescribes the unsteady evolution of the bubble. In this work, we primarily present
 steady solutions, computed by setting $\partial
{\bf R}/\partial t=0$.
However, the time-derivative term in the equations is important for the linear stability calculations presented throughout. 

The dynamic boundary condition (\ref{PressureJump}) is the
non-dimensional form of the Young-Laplace equation, where $p_{b}$ is
the pressure inside the bubble and $\kappa$ is the non-dimensional
curvature of the interface in the $(x,y)$ plane, which we shall term
the in-plane curvature. The other component of curvature, $1/b(y)$,  will be
termed the transverse curvature. Note that in the case $b(y)=1$, the geometry of the channel reduces to a rectangular Hele-Shaw channel, and the equations reduce to those for classic Hele-Shaw flow.
The bubble velocity $U_b$ is an unknown, chosen so that
that the geometric centroid of the bubble is fixed at $x=0$; this velocity may vary in time. The fluid pressure
is fixed to be zero far ahead of the bubble, which means that the bubble
pressure $p_{b}$ is also an unknown with the associated constraint
that the bubble volume must remain a prescribed constant, $V_{0}$,
during the evolution 
  \begin{equation}\label{VolCons}
    \int\hspace{-0.2cm}\int_{\Omega_{b}} b(y) \,\mbox{d}A = V_{0} = \frac{\pi D^2}{4},
  \end{equation} 
  assuming that the bubble occupies the full height of the channel.
  In the model, we neglect the effects of the thin films that
  separate the bubble from the top and bottom boundaries of the
  channel in this volume constraint and also in the two boundary conditions at the bubble interface.
  

The model is solved using the finite element library {\texttt
  oomph-lib} \cite{HeilHazel2006} and implementation details are given
in \cite{Thompson2014,FrancoGomez2016, FrancoGomez2017}. Where needed, linear stability calculations are performed by formulating and solving a generalised eigenvalue problem within the \texttt{oomph-lib} framework, as discussed in some detail in \cite{Thompson2014}.
The centroid position $y_c$ is calculated directly from the bubble boundary, using the same algorithm as for the experimental observations. 

Naturally, the model does not capture all the features of the
three-dimensional system. As mentioned above, our model neglects any thin films above and below the bubbles. A number of formulations have been proposed to correct for the presence of the film, but are generally only appropriate if the film thickness is spatially uniform. These corrections take the form of an increased propagation speed \cite{SaffmanTaylor1958} due to a modified mass balance at the propagating interface, or a reduction of the effective lateral curvature in the normal pressure jump by a factor of $\pi/4$ \cite{ParkHomsy1984, ReineltSaffman1985}. We note that this latter result only applies at small $Ca$, large $\alpha$ and  for a rectangular channel. It is difficult to apply these corrections in our model as we do not operate fully in this asymptotic regime, the presence of the rail means it is unlikely that film thicknesses are spatially uniform, and we do not have a predictive model for these thicknesses. Thus, any such correction would be necessarily ad-hoc. We note that for the case of finger propagation with a similar depth-profile to that explored here, we have previously shown that the uncorrected model is in quantitative agreement with the experimental results when $Ca \lesssim 0.012$, $\alpha_h \le 0.12$ and $\alpha\gtrapprox 40$.

 The low order of the model means that we can satisfy only one boundary condition on each part of the fluid boundary, and we choose to apply the normal pressure jump at the bubble and the no-penetration condition at the side wall. This means that we cannot apply the no-slip boundary condition along the channel side walls, nor the condition of zero tangential stress at the bubble interface. 
It would be possible to apply these boundary condition by including higher order derivatives in the model, such as via the Brinkman equations, but these Brinkman effects are just one of a number of possible corrections (including the thin films suggested above), and we prefer to focus on the uncorrected model presented here. 
Again, we note that quantitative agreement has been reached for air fingers in wide aspect ratio systems without incorporating the tangential stress boundary condition.

We would expect the no-slip and no-tangential stress conditions to be most significant when the bubble boundary is very close to the channel walls  (e.g. the inset image marked (a) in Figure
\ref{fig:Finger_Bubble}b).
In fact, we found that we were unable to compute
asymmetric steady solutions to the model at very low $Ca$; instead each solution branch terminates at a finite value of $Ca$, marked in the figures by a filled circle. This minimum value of $Ca$ does not appear to correspond to a bifurcation point, and the behaviour of the bubbles in this regime is similar to that observed experimentally.


\section{Results}

\subsection{From finger to bubble propagation}
\label{sec1}

\begin{figure}
     \begin{center}
       \includegraphics[width=12.0cm]{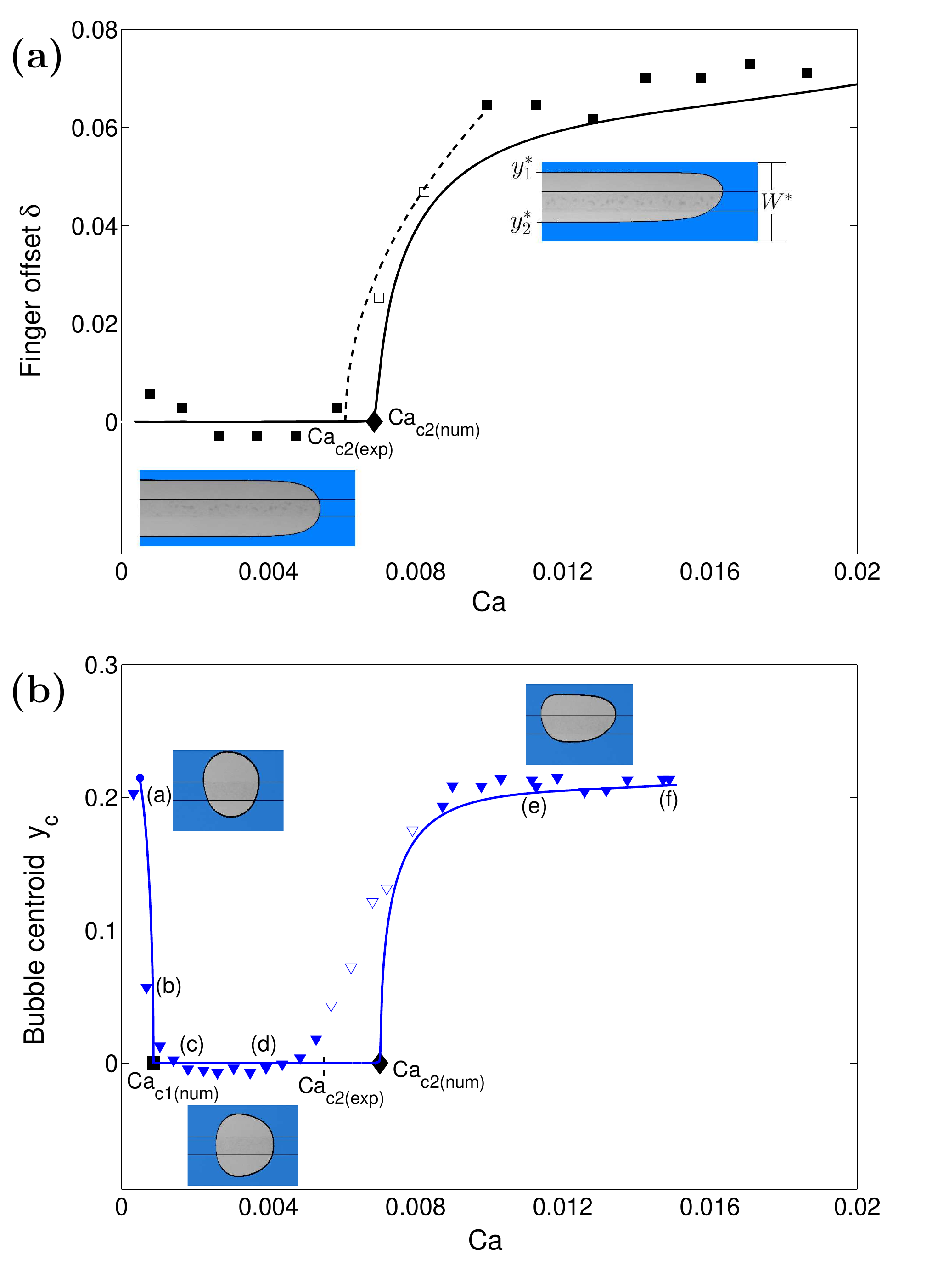}
     \end{center}
     \vspace{-0.7cm}
     \caption{\small Comparison between experimental (symbols) and numerical (solid lines)
       steady modes of propagation in a channel with aspect ratio
       $\alpha=30$, $\alpha_h=0.024$, as a function
       of capillary number $Ca$. 
       The measured rail width is $\alpha_w = 0.233$; for the numerical model, we use  $\alpha_w=0.288$ and sharpness
       parameter $s=40$. Filled symbols
       correspond to steady modes of propagation and open symbols
       denote modes of propagation that do not reach a steady state
       within the length of the channel. (a) Finger propagation, where
       the finger offset is defined as $\delta = 
         (y^*_1+y^*_2)/(2 W^*)$; for large bubbles $2\delta \approx y_c$ where $y_c$ is the centroid offset. The symmetric finger loses
       stability to an asymmetric finger through a supercritical
       pitchfork bifurcation. The experimental critical capillary
       number, $Ca_{\rm c2(exp)}$, was determined by fitting a
       square-root function to data points just past the bifurcation
       point (dashed line) \cite{DeLozar2009}. (b) Propagation of a
       bubble of effective static diameter $D=0.76$ ($D^*=22.8$
       mm). The black symbols denote the locations of supercritical
       pitchfork bifurcations.  The filled circle on the 
       computed stable propagation modes
       indicates the smallest value of $Ca$ for which a solution
       could be computed.}\label{fig:Finger_Bubble}
   \end{figure}

We begin by presenting new
results for the propagation of a finger (a semi-infinite
bubble), in the presence of the rail of modest height used in this
study ($\alpha_h=0.024$, i.e. 2.4\% of the height of the
channel). Steady modes of finger propagation are quantified in Figure
\ref{fig:Finger_Bubble}a  in terms of the offset $\delta$ of the
finger tip from the axial centreline of the channel (see figure caption for definition of $\delta$) as a function of
the capillary number \cite{FrancoGomez2016}. The finger is symmetric about the
centreline of the channel in the capillary-static limit, and this
symmetric finger propagates stably as $Ca$ is increased  up to a
critical value of the capillary number, $Ca_{\rm c2(exp)}  \approx 6.1 \times
10^{-3}$, at which point the finger undergoes a supercritical symmetry-breaking
bifurcation to an asymmetric finger. The unavoidable small inherent bias
present in the experimental channel promotes asymmetric propagation
primarily on one side of the rail. The open symbols denote fingers
that were still transiently evolving towards an off-centre steady state when they
reached the end of the experimental channel. These points, which are
typically clustered just past the bifurcation point, reflect the
critical slowing down of finger evolution near the bifurcation
point. Transient states were not observed for $Ca<Ca_{\rm c2(exp)}$
because both the initial position of the finger and the stable steady state were centred. 

Steady numerical solutions of the depth-averaged model are shown in
Figure \ref{fig:Finger_Bubble}a by a solid line. The results are qualitatively similar to those previously presented
by \cite{FrancoGomez2016}, but in that work $\alpha_w$ was
20\% smaller. Comparison of the results for the two rail widths in \ref{A2}  indicates that the narrower rail is associated with a slightly larger value of the critical capillary number $Ca_{\rm c2(num)}$. The numerical
results are in good agreement with the experiment, although the
bifurcation point occurs for $Ca_{\rm c2(num)} \simeq 6.8\times
10^{-3}$, which is larger than $Ca_{\rm c2(exp)}$ by approximately
11\%. This result, obtained for $\alpha =30$, is consistent with the
findings of Franco-Gomez et al. \cite{FrancoGomez2016}, who
demonstrated quantitative agreement between experiment and model, with
less than 5\% discrepancy, for aspect ratios $\alpha \ge 40$. However,
for $\alpha =20$, $Ca_{\rm c2(num)}$ exceeded $Ca_{\rm c2(exp)}$ by
approximately 50\%. This dependence on $\alpha$ is consistent with
the model assumptions; the validity of the
two-dimensional model relies on sufficiently high aspect ratio
channels, in addition to small occlusion heights and small values of
$Ca$.

The loss of symmetry of the propagating mode is due to the
variation of local viscous resistance across the channel due to the presence
of the rail.  For a near-uniform axial pressure gradient, both fluid
and interface will propagate faster in the off-rail regions, but this
speed differential is resisted by surface tension acting through the
in-plane curvature of the finger to stabilise the finger in a centred,
symmetric configuration over the rail. For
sufficiently large flow rates, the viscous forces dominate and
variations in the local interface speed near the finger tip are able to drive
the finger into the off-rail regions, resulting in
the observed symmetry-breaking of the finger at
a critical value of $Ca$ \cite{Thompson2014,FrancoGomez2016}. 

 In Figure \ref{fig:Finger_Bubble}b, we present the steady propagation modes of a finite, but relatively large bubble, started from a centred initial position with $D=0.76$, where
the centroid of the bubble is plotted as a function of
$Ca$. Analogously to the finger case shown in Figure
\ref{fig:Finger_Bubble}a, the bubble loses symmetry through a
supercritical pitchfork bifurcation ({\large {$\blacklozenge$}}),
at similar values of $Ca_{\rm c2(exp)}$ and $Ca_{\rm  c2(num)}$ to the
finger. The off-centre propagation of both bubbles and fingers at high flow rates is supported by the reduced viscous resistance in the deeper parts of the channel, which dominates over capillary effects when $Ca$ is large. However, a key
difference in the bubble scenario
is the absence of a symmetric capillary-static state. The
bubble can only adopt a stable symmetric capillary-static configuration once its 
volume is sufficient that it spans the entire width of the
channel \cite{Hazel2013}. When the
volume of the bubble is reduced below this threshold, the symmetric
capillary-static state is no longer stable to lateral perturbations because
the reduction in transverse curvature when the bubble expands into one of
the deeper side channels on either side of the rail drives a flow
that pushes the bubble further to the side until an asymmetric equilibrium
is reached; see Section \ref{sec2} for further discussion. 
The introduction of axial flow, however, changes the behaviour: for
small values of $Ca$, for which variations in viscous
resistance across the channel are small compared to surface tension, the bubble inclines so that
its tip is nearer the centreline, which acts to drive the bubble
towards an on-rail position \cite{FrancoGomez2017}. Thus, the
capillary-static induced flow and viscous-pressure-driven flows act in
opposite directions and for moderately large flow rates, the centred
mode of propagation is stabilised, as shown in Figure
\ref{fig:Finger_Bubble}b. The flow-induced stabilisation of this
symmetric mode takes place through a supercritical pitchfork
bifurcation  (\textcolor{black}{$\blacksquare$}) at $Ca_{\rm
  c1(num)} < Ca_{\rm c2(num)}$, which is in good agreement with the experimental
value. Hence, the bubble propagates symmetrically about the centreline
of the channel over a limited range of $Ca$ between the two
bifurcations. 
Note that near the bifurcation, the experimental data for bubbles is less well approximated by a square-root function  than the results for fingers, suggesting that the bubbles are more sensitive to imperfections. 

  \begin{figure}
    \begin{center}
       \includegraphics[width=\textwidth]{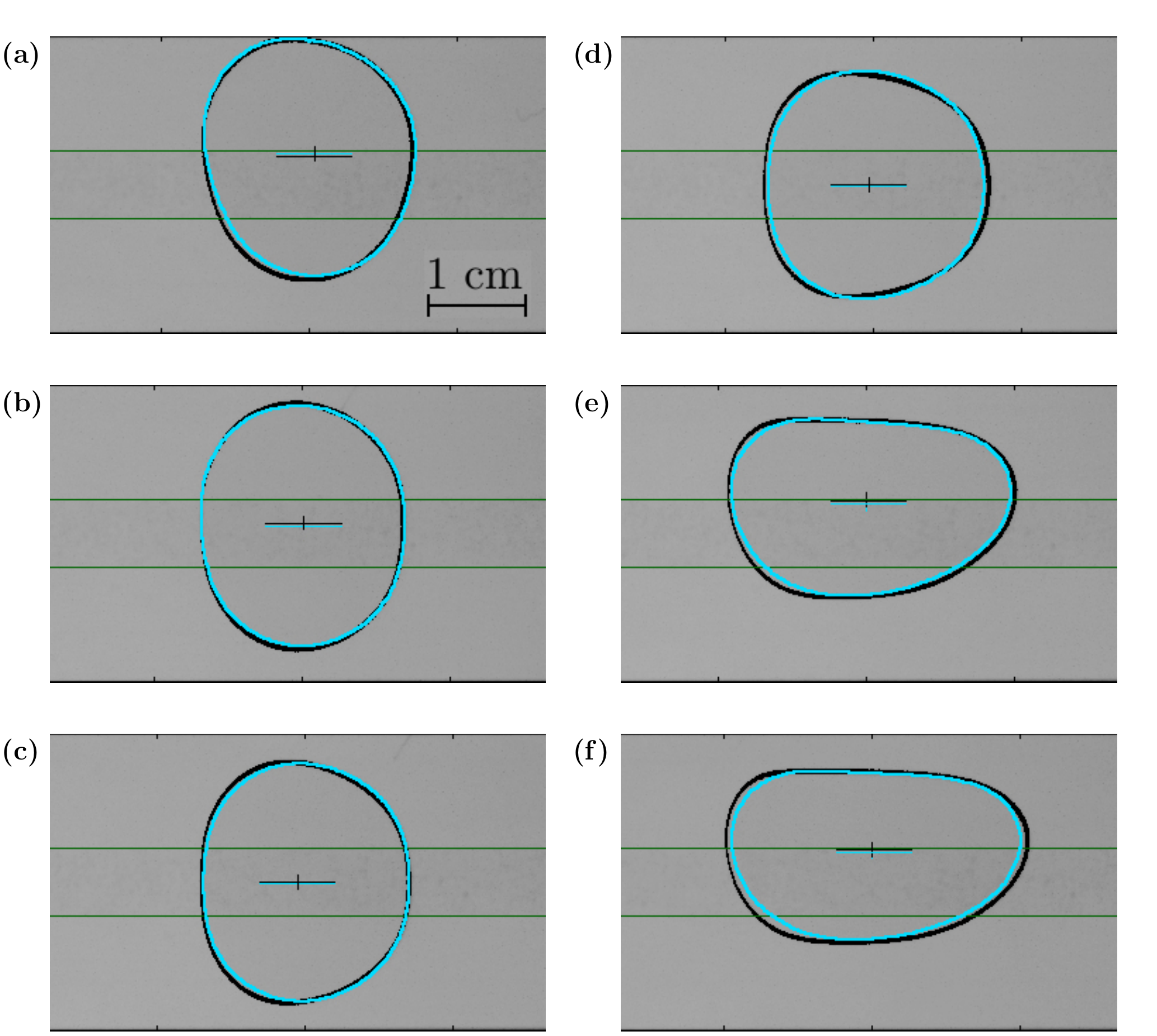}
     \end{center} 
     \caption{\small Direct comparison between experimental and
       numerical bubble outlines for increasing values of $Ca$, 
       bubbles with static effective diameter $D=0.76$ ($D^*=22.8$ mm). 
       The computed bubbles shapes are shown
       with solid cyan lines, experimental bubble outlines 
       with solid black lines, and the edge of the rail is highlighted
       with a solid green line. The capillary numbers (dimensional
       flow rates) are: (a) $Ca_{num}=5.34\times 10^{-4}$, $Ca_{exp} = 3.26\times 10^{-4}$ ($Q^*=1$ ml/min);
       (b) $Ca_{num}=7.74\times 10^{-4}$, $Ca_{exp} = 6.80\times 10^{-4}$ ($Q^*=2$ ml/min); (c)
       $Ca=1.81\times 10^{-3}$ ($Q^*=5$ ml/min); (d) $Ca=3.92\times
       10^{-3}$ ($Q^*=10$ ml/min); (e) $Ca=1.12\times 10^{-2}$
       ($Q^*=23$ ml/min); and (f) $Ca=1.49\times 10^{-2}$ ($Q^*=30$
       ml/min). Note that in (a) and (b) only, the numerical value of $Ca$ is chosen so that the shapes have similar offset values; otherwise the same value of $Ca$ is used for both experimental and numerical results.
}\label{fig:BubbleShapeCompQ}
  \end{figure}

 Figure \ref{fig:BubbleShapeCompQ} shows a direct comparison of
 experimental and numerical bubble shapes corresponding to the data
 shown in Figure \ref{fig:Finger_Bubble}b, with the numerical bubbles shown by cyan outlines overlaid on experimental images where the bubble boundary is visible as a black line. 
 The scale for the images  is chosen to match the channel width, with the downstream location chosen to match the axial position of the centroid.  Both the lateral components of the 
 centroids and the overall shapes of the
 experimental and numerical bubbles outlines are in good
 agreement for the three observed propagation modes: asymmetric low-$Ca$,
 Figure \ref{fig:BubbleShapeCompQ}a,b; symmetric, Figures
 \ref{fig:BubbleShapeCompQ}c,d; and asymmetric high-$Ca$, Figures
 \ref{fig:BubbleShapeCompQ}e,f.
 
 Although the bubble  shapes are in good agreement, the projected area of the experimental
 bubble is consistently larger than that of the numerical bubble, and
 this discrepancy increases with increasing $Ca$. 
  As $Ca$ increases in the experiment, the thickness of the liquid
 films separating the bubble from the top and bottom
 boundaries of the channel increases
 \cite{ReineltSaffman1985, FrancoGomez2017}, which in turn enlarges
 the projected area of the experimental bubble. The difference in
 bubble projected areas between experiment and model arises because the model does not account for
 the presence of these liquid films, instead assuming that the bubble always occupies the whole height of the channel for the purpose of applying the volume constraint. The experimental projected area
 exceeds the numerical one by only 2.6\% in  Figure
 \ref{fig:BubbleShapeCompQ}a, but this value rises to 10.4\% in
 Figure \ref{fig:BubbleShapeCompQ}f. We refer to \ref{A3} for
 systematic measurements of bubble area variations with $Ca$ over the
 entire range of $Ca$ investigated. 

\subsection{Effect of bubble size on the steady modes of bubble propagation}
\label{sec2}

 \begin{figure}[ht]
     \begin{center}
       \includegraphics[width=12.0cm]{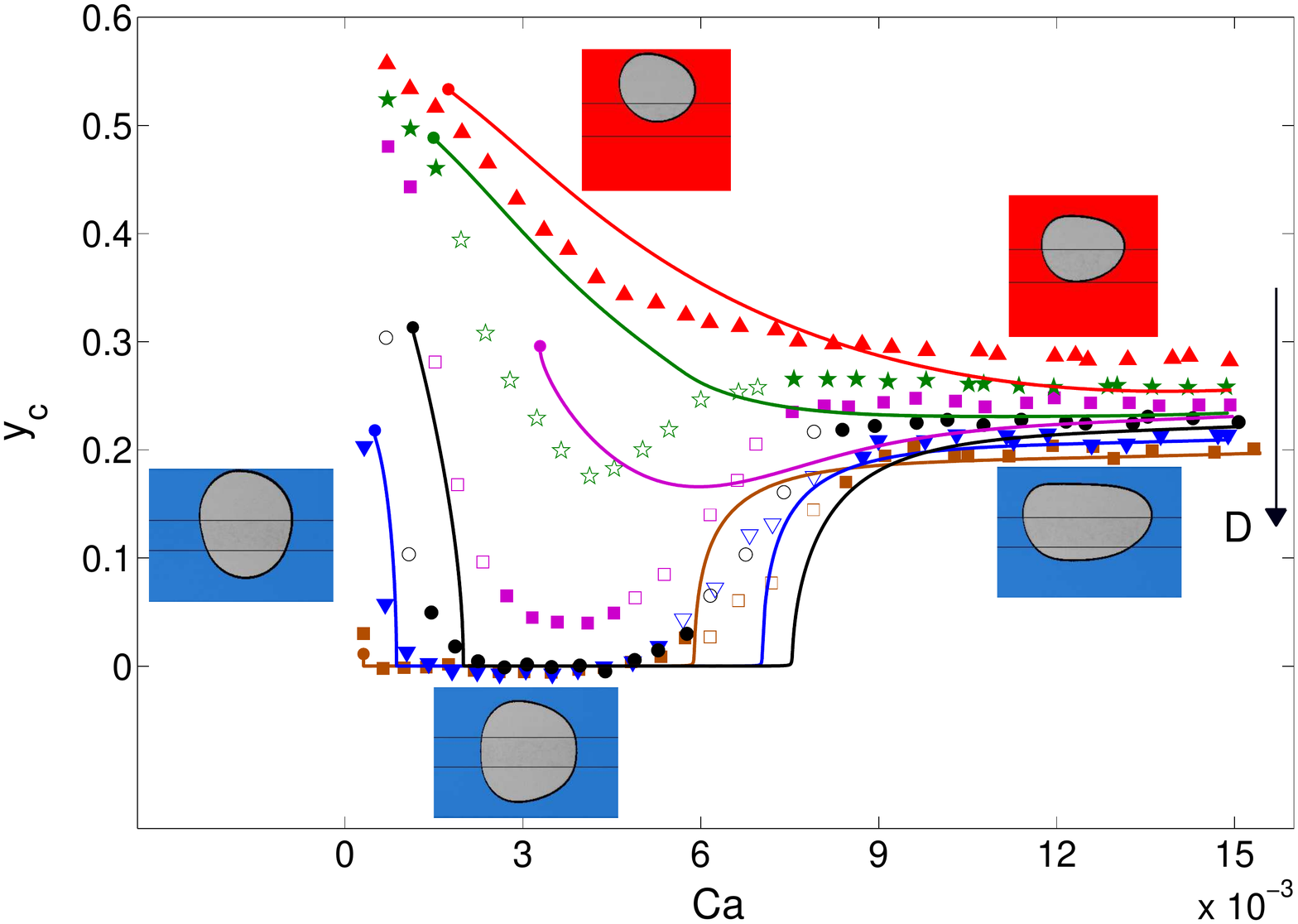}
     \end{center}
     \vspace{-0.5cm}
     \caption{\small Comparison between experimental and numerical
       bubble centroid displacements from the centreline, $y_c$, as a
       function of $Ca$, for different bubble sizes quantified by
       their static non-dimensional (dimensional) diameters:
       (\textcolor{red}{\large$\blacktriangle$}) $D=0.50$ ($D^*=14.9$
       mm); 
       (\textcolor{darkgreen}{\bf$\bigstar$}) $D=0.55$
       ($D^*=16.5$ mm); 
       (\textcolor{magenta}{$\blacksquare$}) $D=0.59$
       ($D^*=17.8$ mm);  
       (\textcolor{black}{\Large\bf$\bullet$})
       $D=0.67$ ($D^*=20.1$ mm);
       (\textcolor{blue}{\large$\blacktriangledown$}) $D=0.76$
       ($D^*=22.8$ mm) and 
       (\textcolor{darkbrown}{$\blacksquare$}) $D=0.87$
       ($D^*=26.1$ mm).  Each data point corresponds to a bubble
       propagating under constant imposed flow rate from an initially
       centred position. The numerically determined steady states corresponding to
       these bubble sizes are shown with solid lines of matching colour. Note that for small values of $Ca$ the curves are shifted monotonically towards smaller values of $y_c$ with increasing bubble diameter. The filled
       circle marking the lower $Ca$ extremity of each line correspond
       to the smallest value of $Ca$ for which a solution could be
       computed. The channel parameters are $\alpha=30$,
       $\alpha_h=0.024$,  $\alpha_w=0.288$ and $s=40$. 
       }\label{fig:ExpNumFingerOffset_lowCa}
   \end{figure}

Having established the bifurcation sequence for varying $Ca$ that underpins the
propagation of finite bubbles from a centred initial position, for a single moderately sized bubble,
we show the evolution of this picture for decreasing
bubble size in Figure \ref{fig:ExpNumFingerOffset_lowCa}.  For very small
values of $Ca$, the displacement of the bubble centroid from the
centreline increases significantly as the bubble size is reduced,
suggesting increasingly asymmetric capillary-static
configurations. This is simply because a greater proportion of a
smaller bubble fits within the off-rail region.
In addition, smaller bubbles have larger in-plane
curvatures, and thus the surface-tension-induced pressure
variations are greater than for larger bubbles. 
Hence, larger viscous forces are in turn required in order to
stabilise symmetric modes of propagation. This means that the critical
capillary number associated with the first supercritical pitchfork
bifurcation, which stabilises the steady symmetric bubble, increases
as the bubble size is reduced. In fact, the computed values of
$Ca_{\rm c1(num)}$ for the cases shown in Figure \ref{fig:ExpNumFingerOffset_lowCa} closely match the experimental values of $Ca_{\rm
  c1(exp)}$. Moreover the centroid positions of the asymmetric bubbles
at very low values of $Ca$ below $Ca_{\rm c1}$ remain well captured by
the numerical computations. The experimental data at small $Ca$ for
the three largest bubbles indicate that $Ca_{\rm c1}$ decreases as the
bubble size increases, suggesting that a capillary-static
symmetric bubble would eventually be recovered if the bubble size was further
increased, as expected based on the behaviour of semi-infinite air fingers.

\begin{figure}
    \begin{center}
              \includegraphics[width=16.0cm]{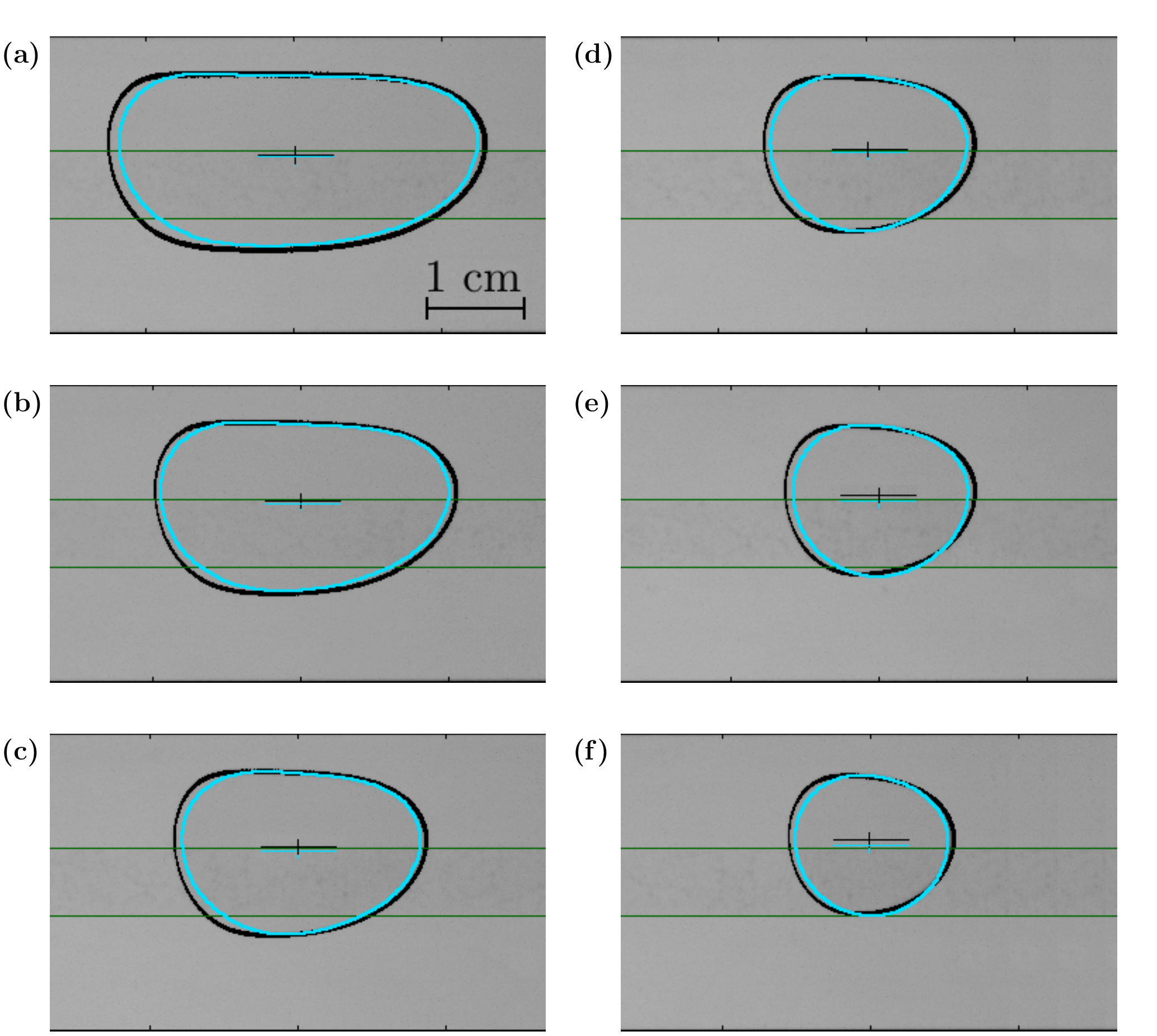}
     \end{center} 
     \caption{\small Direct comparison between experimental and
       numerical bubble outlines for decreasing bubble sizes. The
       bubbles were propagated  with the highest values of
       $Ca\simeq1.5\times 10^{-2}$ ($Q^*=30$ ml/min) shown in  Figure
       \ref{fig:ExpNumFingerOffset_lowCa}. The computed bubbles shapes
       are shown with solid cyan lines, experimental bubble outlines
       are plotted with solid black lines, and the edge of the rail is
       highlighted with a solid green line. The bubble sizes are
       quantified by the effective nondimensional (dimensional)
       diameters of the static bubbles: (a) $D=0.87$ ($D^*=26.1$ mm);
       (b) $D=0.76$ ($D^*=22.8$ mm); (c) $D=0.67$ ($D^*=20.1$ mm);
       (d) $D=0.59$ ($D^*=17.8$ mm); (e) $D=0.55$ ($D^*=16.5$ mm)
       and (f) $D=0.50$ ($D^*=14.9$
       mm).
}\label{fig:BubbleShapeCompArea}
  \end{figure}
 
The three smaller bubbles in Figure \ref{fig:ExpNumFingerOffset_lowCa}
do not exhibit symmetric propagation at any $Ca$. This is because the
viscous forces required to compensate for the destabilising effect of
surface tension forces become sufficiently large that they induce
significant variations in flow across the channel.  In response
to a perturbation of the bubble, the flow variations have
the effect of tilting the bubble tip away from the rail and into the
deeper side channel, so that stabilisation of the symmetric mode
of propagation is prevented \cite{FrancoGomez2017}. The disappearance
of the stable symmetric bubble occurs through the coalescence of the
two nearby pitchfork bifurcations for a critical bubble diameter $D_c$, in the region
$0.59<D_c<0.67$ (i.e. between the magenta and black symbols in Figure \ref{fig:ExpNumFingerOffset_lowCa}). For bubbles below this critical diameter, the
bubble always propagates in an off-centred position, but there is still a noticeable shift at intermediate $Ca$ towards the centre of the channel due to the stabilising effect of viscous forces. When
$Ca$ becomes sufficiently large, $y_c$ increases again and
eventually reaches an approximately constant value for sufficiently
large $Ca$. The centroid positions $y_c$ of the experimental asymmetric
states on the right-hand side of Figure
\ref{fig:ExpNumFingerOffset_lowCa} are accurately captured by the
model for the three largest bubbles. However, the smaller three
bubbles exhibit discrepancies in $y_c$ at these large $Ca$, most likely due to enhanced three-dimensional flow features
for reduced bubble aspect ratios.

The coalescence of the two pitchforks leads to a stable asymmetric branch for both the experimental and numerical results shown in Figure \ref{fig:ExpNumFingerOffset_lowCa}. However, these two branches are not in quantitative agreement.
This is in part due to the
sensitivity of the system to small bubble size changes in this
regime, but also because of the long duration of transient
evolution of the propagation modes, which means that steady states are
not typically reached within the length of the experimental
channel. These long transients, together with 
unavoidable imperfections in the experimental
system, mean that the bifurcation regions for the three larger bubbles
differ between the
model and experiments, particularly near $Ca_{\rm c2}$. In the model
calculations the location of $Ca_{\rm c2(num)}$ varies
non-monotonically with bubble size, which is due to the
subtle and non-trivial
nature of the interplay between viscous and surface tension
forces that leads to destabilisation of the symmetric solution. 
It is not possible to determine any similar trends for the
experimental data which all agree, within experimental error,
for the three largest bubbles in the region near $Ca_{\rm  c2(exp)}$.

A direct comparison between experimental and numerical bubble outlines
for the bubbles sizes shown in Figure \ref{fig:ExpNumFingerOffset_lowCa}
is presented in Figure \ref{fig:BubbleShapeCompArea} at the largest
value of $Ca$ considered, $Ca\simeq1.5\times 10^{-2}$.
As was the case in Figure \ref{fig:BubbleShapeCompQ}, the presence of thin liquid films above and below the
bubble lead to the experimental projected area exceeding
that of the numerical bubble. The two areas differ by 
 an approximately constant factor of
$10.6 \pm 0.9 \%$.  In \ref{A3}, we track bubble area across the range of $Ca$ and $D$ used in Figure \ref{fig:ExpNumFingerOffset_lowCa}, and find that the area ratio increases with $Ca$, but is largely independent of bubble size. This implies that the average film
 thickness is independent of bubble size at fixed capillary number.
Despite this discrepancy in area, however, the numerical
bubble shapes in Figure \ref{fig:BubbleShapeCompArea} closely match those for the experimental bubbles, and the crosses
denoting the centroid positions of the numerical and experimental
bubbles highlight the small differences in $y_c$. As 
previously discussed in relation to Figure
\ref{fig:ExpNumFingerOffset_lowCa}, at large $Ca$ the discrepancies in $y_c$ increase only modestly as the
bubble size is decreased.

\subsection{Numerical bifurcation diagram}\label{sec3}

 \begin{figure}
    \begin{center}
  \includegraphics[width=13cm]{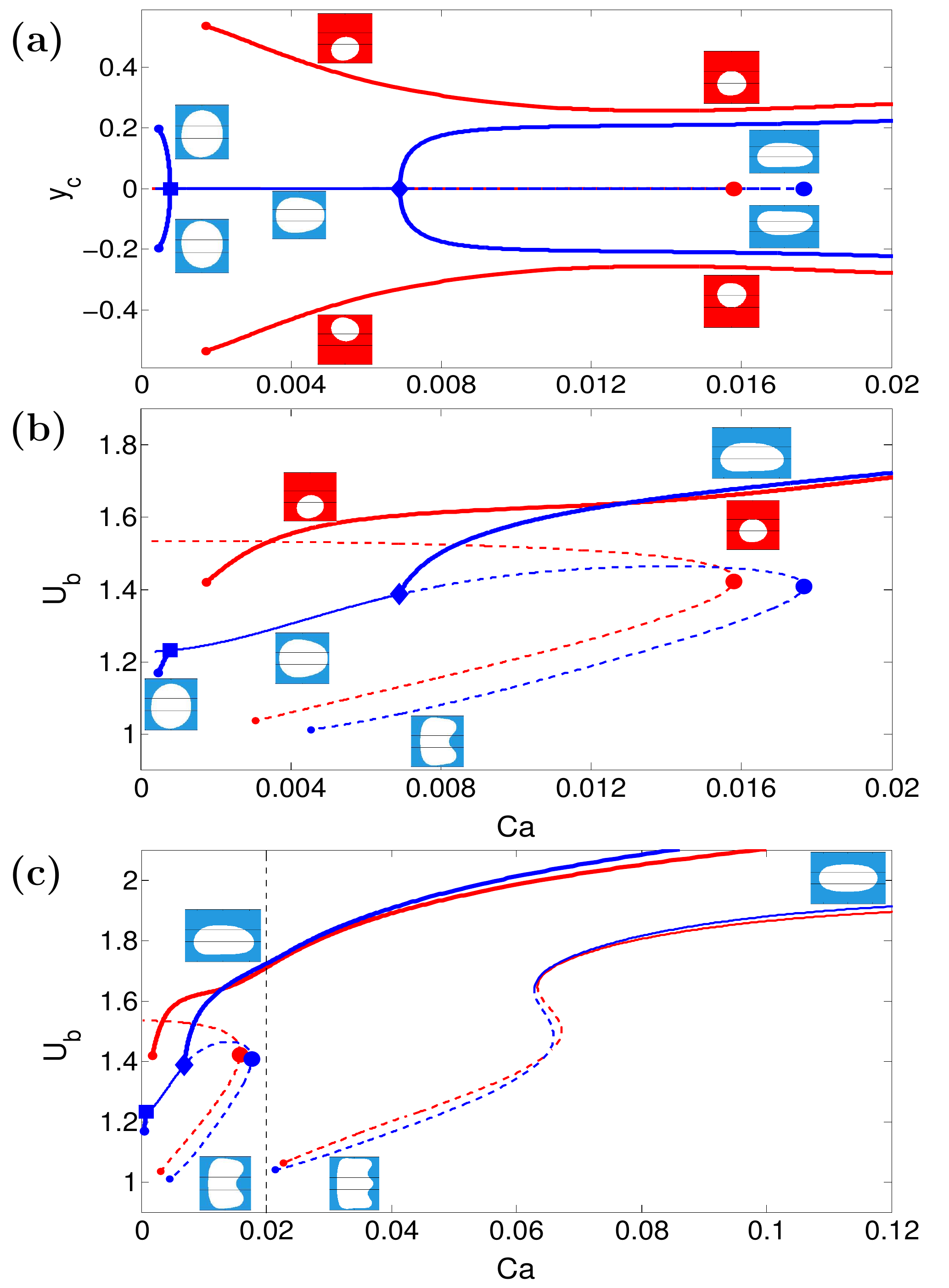}
     \end{center}
     \vspace{-0.5cm}
     \caption{\small Numerical simulations of (a) centroid position $y_c$ and (b,c) nondimensional
       bubble speed $U_b$ as a function of capillary number $Ca$ for
       bubble sizes $D=0.5$ ($D^*=15$ mm) in red and $D=0.77$
       ($D^*=23.2$ mm) in blue. Solid thick lines correspond to
       asymmetric stable states, solid thin lines to symmetric stable
       states and dashed thin lines to symmetric unstable states
       The channel parameters are $\alpha=30$,
       $\alpha_h=0.024$,  $\alpha_w=0.288$ and $s=40$.
       Projection (a) shows the pitchfork bifurcation to off-centre branches, but there are also multiple centred branches, which are revealed in projection (b). At larger $Ca$ in (c), there is a disconnected symmetric branch including a family of stable propagation modes and hence both symmetric and asymmetric propagation modes are stable for $Ca\gtrapprox 0.06$. \label{fig:Bif_UlowCa_DD_TT}
     }
  \end{figure}

In addition to predicting the stable modes of steady propagation of the
bubble, the numerical simulations can also capture unstable solutions.
Steadily propagating bubbles are quantified in Figure
\ref{fig:Bif_UlowCa_DD_TT} in terms of their centroid position (a) and non-dimensional
propagation speeds as functions of $Ca$ (b,c). While the two asymmetric branches emerging from
each pitchfork bifurcation are differentiated in projection (a), they collapse onto a single line in projection (b). Stable
(unstable) modes of propagation are shown with solid (dashed) lines,
respectively. The bifurcation diagram shown in Figure
\ref{fig:Bif_UlowCa_DD_TT}a spans a similar range of capillary
numbers to that studied in section \ref{sec2}. The blue line corresponds to the same large
bubble ($D=0.77$) as in Figures \ref{fig:Finger_Bubble}b and \ref{fig:BubbleShapeCompQ} and is also shown in blue in Figure \ref{fig:ExpNumFingerOffset_lowCa}. At very low values of $Ca$,  this large bubble propagates stably in an asymmetric configuration. The asymmetric solution branch disappears at
the first supercritical pitchfork bifurcation point
(\textcolor{blue}{$\blacksquare$}), enabling stable symmetric bubbles to
propagate as $Ca$ is increased further until the second supercritical
pitchfork bifurcation point ({\color{blue}$\Diamondblack$}), where
the symmetric mode loses stability in favour of an asymmetric mode, as
discussed in section \ref{sec1}. In contrast, the smaller bubble (red
lines, $D=0.5$) propagates stably in an asymmetric configuration over the
entire range of $Ca$ investigated numerically. In this case, the
symmetric branch is disconnected from the asymmetric branch and
remains unstable for all values of $Ca$. 

For both bubbles, the
unstable symmetric branch undergoes a saddle-node bifurcation at a
limit point to an unstable dimpled bubble solution, the first
Romero--Vanden-Broeck (RVB) solution, which was originally uncovered in the
context of Saffman--Taylor fingering \cite{Romero1982,
  VandenBroeck1983,GardinerMcCue2015}.
In Figure \ref{fig:Bif_UlowCa_DD_TT}c, we
increase the range of capillary numbers by a factor of six, and
capture the second RVB solution, which exhibits two dimples at its tip
but is also unstable. Green \textsl{et al.} \cite{Greenetal2017} have
recently investigated bubble shapes in unbounded Hele-Shaw flow and
have found a countably infinite number of disconnected solution branches, each
corresponding to a particular number of dimples at the bubble tip. At fixed dimensionless speed $U_b = U^{*}_{b}/U^{*}_{0}$, the branches occur at increasing $Ca$ for increasing dimple number.
We expect similar solution branches to exist in our system, despite the presence of the bounding channel walls and the rail, but we have not attempted to compute them.
With increasing $Ca$, the second RVB solution in Figure \ref{fig:Bif_UlowCa_DD_TT}c undergoes two
saddle-node bifurcations, while its shape evolves towards a round
tipped bubble. These bifurcations mean that a symmetric mode of propagation is restabilised
at large values of $Ca$, analogously to the previously studied
finger propagation \cite{FrancoGomez2016}. In experiments, the
observed steady propagation modes should be stable,  but we show in
section \ref{sec4} that unstable solutions do arise transiently.

\subsection{Unsteady bubble propagation at large flow rates} 
\label{sec4}

 \begin{figure}[ht]
     \begin{center}
       \includegraphics[width=16.0cm]{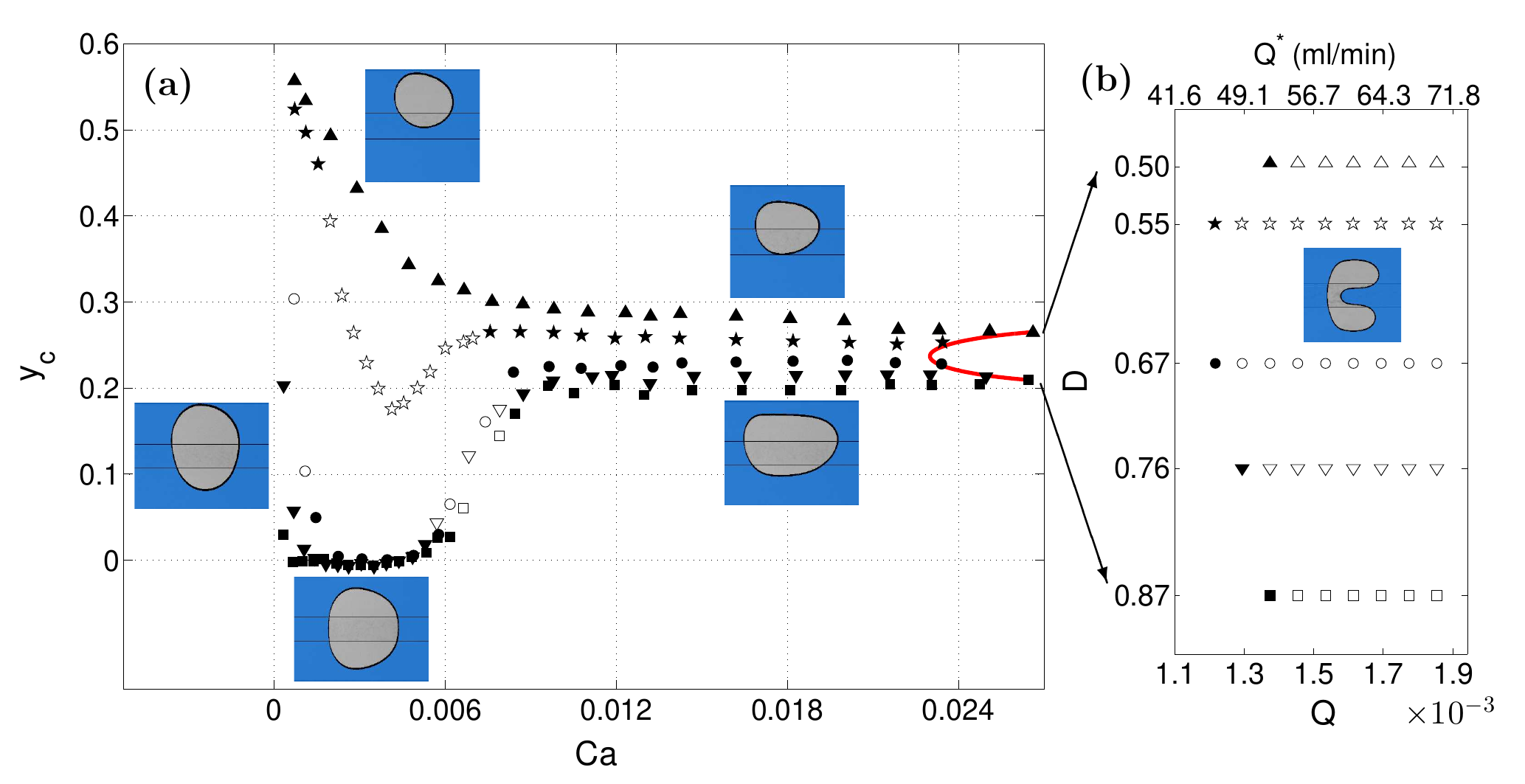}
     \end{center}
     \vspace{-0.5cm}
     \caption{\small (a) Experimental centroid position of a steadily
       propagating bubble, started from a centred initial condition,
       as a function of capillary number for the same channel geometry
       but an extended range of $Ca$ compared to Figure
       \ref{fig:ExpNumFingerOffset_lowCa}:
       (\textcolor{black}{\large$\blacktriangle$}) $D=0.50$
       ($D^*=14.9$ mm), (\textcolor{black}{\bf$\bigstar$}) $0.55$
       ($16.5$ mm), (\textcolor{black}{\Large\bf$\bullet$}) $0.67$
       ($20.1$ mm), (\textcolor{black}{\large$\blacktriangledown$})
       $0.76$ ($22.8$ mm) and (\textcolor{black}{$\blacksquare$})
       $0.87$ ($26.1$ mm). The red line indicates the maximum values
       of $Ca$ for which bubbles of different sizes propagate
       steadily. (b) Values of non-dimensional (dimensional) flow
       rates for which time-evolving dimpled bubbles are observed
       (open symbols) or remain convex (closed symbols) for five different bubble
       sizes. }\label{fig:ExpFingerOffset_lowCa}
   \end{figure}

Thus far, we have focused on a range of bubble sizes and capillary
numbers for which an initially centred bubble rapidly evolves towards
a stable steady mode of propagation, which may or may not be symmetric about the centreline of the channel. The second pitchfork
bifurcation occurs at $Ca_{\rm c2(exp)} \simeq 6 \times 10^{-3}$ as
shown in Figure \ref{fig:ExpNumFingerOffset_lowCa}, and beyond this
point the only steady modes of propagation we have observed experimentally are asymmetric. We
also found numerically that a steady symmetric mode of propagation is
stabilised above the much larger value $Ca \approx 9 \times
10^{-2}$ (Figure \ref{fig:Bif_UlowCa_DD_TT}). Hence, at these values
of $Ca$ the model predicts that the system is bistable and
the natural question now arises: which mode of propagation is selected
at large $Ca$ and does this change for different initial conditions?

Initially asymmetric bubbles driven with a
large $Ca$ always rapidly evolved towards a steadily propagating
asymmetric configuration. However, when the bubble was initially
centred, we found that the bubble did not converge onto a stable mode
of propagation, but instead continually evolved in time until it
broke up into two bubbles. This is illustrated in Figure
\ref{fig:ExpFingerOffset_lowCa}a, where the range of capillary numbers
is extended compared to Figure
\ref{fig:ExpNumFingerOffset_lowCa}. 
The red line in Figure \ref{fig:ExpFingerOffset_lowCa}a denotes the largest
value of $Ca$ for which a steadily propagating asymmetric bubble is
reached after starting from centred initial conditions;
this boundary depends non-monotonically on bubble size.
For larger flow rates, as in Figure \ref{fig:ExpFingerOffset_lowCa}b, the bubble moves into a double tipped configuration, and then breaks up. Note that the dimensionless imposed flow rate $Q = Ca/U_b$ is
used to characterise the unsteady bubbles instead of $Ca$ because bubbles do not necessarily propagate with a constant speed. 

\begin{figure}[ht]
     \begin{center}
       \includegraphics[width=16cm]{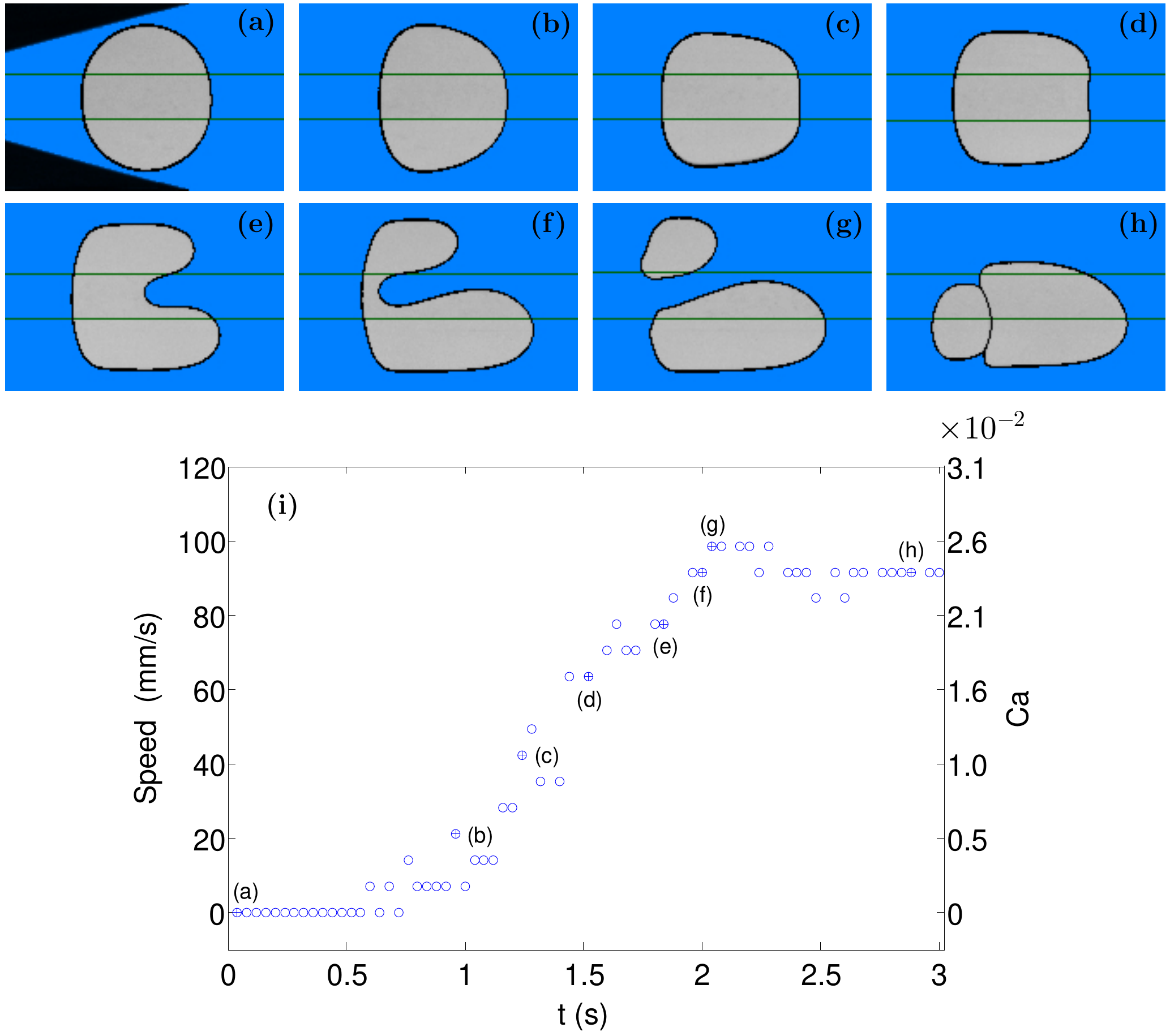}
     \end{center}
     \vspace{-0.5cm}
     \caption{\small (a--h) Evolution of an unsteady bubble with
       diameter $D=0.77$ ($D^*=23.15$ mm) propagating under a large
       flow rate $Q^*=100$ ml/min, from a centred initial
       condition. The channel geometry is the same as in Figure
       \ref{fig:ExpNumFingerOffset_lowCa}. The plot, which shows the
       instantaneous speed of the bubble tip ($Ca$) as a function of time, with
       labels (a--h) referring to the snapshots, indicates that the
       bubble eventually reaches a steady state of changed
       topology.
       }\label{fig:DoubleTippedBubble_lowCa}
   \end{figure}

The evolution of a bubble before and after break-up is illustrated in Figure
\ref{fig:DoubleTippedBubble_lowCa}, where its instantaneous shape and speed are
shown as a function of time. When the bubble is set into motion from a
centred position (a), it rapidly expands across the channel  (b, c)
and its front deforms over the rail to form an approximately centred
dimple (d), because of the increased local viscous resistance above the centred rail. The bubble tips thus formed on either side of this dimple are
displaced at a faster speed than the central region of the front
(e, f). Typically, one tip develops more rapidly than the other, so
that the bubble evolves asymmetrically about the centreline of the
channel. If the asymmetry remains moderate, the growth of the two tips
results in the break-up of the bubble into two smaller bubbles (g),
which aggregate to reach a steady mode of propagation (h). The
evolution of the bubble is reflected in the plot of instantaneous
speed as a function of time, which shows a monotonic increase towards
a final steady  propagation speed.  

In terms of nonlinear dynamics, this means that the centred initial
bubble does not lie within the basin of attraction of steady modes of
propagation for sufficiently large imposed flow rates and its dynamics
is affected by the presence of unstable solutions. The
first RVB solution is weakly
unstable in the sense that only two of the associated eigenvalues are
unstable (repelling), based on a linear stability analysis performed
when computing the results for Figure \ref{fig:Bif_UlowCa_DD_TT};
the other eigenvalues are stable (attracting) and locally
span a (high-dimensional) space known as the stable manifold. Thus, once
set into motion, we conjecture that the evolving bubble is transiently
attracted towards the vicinity of the stable
manifold of the first RVB solution described in
section \ref{sec3} (dimpled front).

\begin{figure}[ht]
     \begin{center}
       \includegraphics[width=16.0cm]{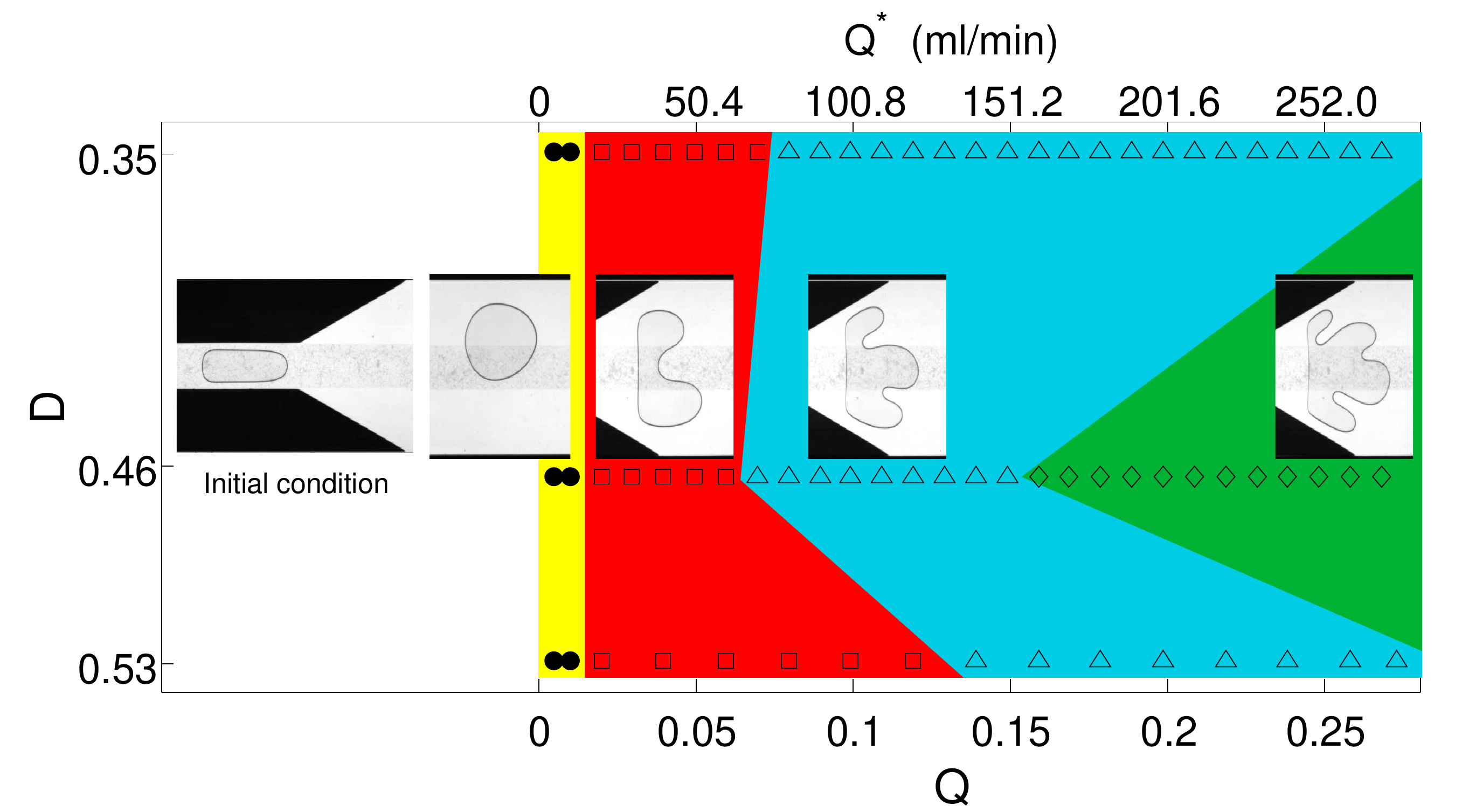}
     \end{center}
     \vspace{-0.5cm}
     \caption{\small Experimental modes of propagation of an initially
       centred bubble in the parameter plane spanning bubble diameter
       and flow rate. Solid (open) symbols represent steady
       (unsteady) propagation: ($\bullet$) steady asymmetric bubble;
       ($\Box$) double-tipped bubble, ($\bigtriangleup$) triple-tipped
       bubble, ({\large$\Diamond$}) quadruple-tipped bubble. Note that
       the channel used for these experiment has an aspect ratio of
       $\alpha=40$, rail width $w=10.0\pm0.1$ mm, and height
       $h=24\pm1$ $\mu$m, and that the silicone oil is more viscous than previously, with $\mu^* = 5.0\times10^{-2}$ Pa s. The coloured
regions are indicative only and used to distinguish
different modes of steady (solid symbols) or unsteady
(open symbols) modes of propagation.}\label{fig:mult_tipbubbles}
  \end{figure}

As discussed in section \ref{sec3}, the first RVB solution is
part of an infinite countable family of bubbles solutions parametrised
by the number of dimples exhibited by their front
\cite{GardinerMcCue2015,Greenetal2017}. Thus, our next question is: could the
initially centred bubble transiently explore other unstable
RVB states? This hypothesis was explored for a slightly larger channel
aspect ratio $\alpha=40$ and rail width $w^*=10.0$~mm. The results
are shown in Figure \ref{fig:mult_tipbubbles} in the parameter plane
spanning bubble diameter and imposed flow rate. 
The fluid viscosity has been increased to enable larger dimensionless flow rates (up to 3.5 times larger than in Figure \ref{fig:ExpFingerOffset_lowCa}) while maintaining good visualisation given limitations on frame rate,  and two of the
three bubbles tested are smaller than previously shown.   For moderate values of $Q$ the
bubble propagates steadily in an asymmetric configuration. However, as
$Q$ is increased, bubble configurations with one, two or three dimples
are successively selected, which suggests that the bubble also
transiently explores the second and third RVB solutions, before
breaking-up to reach a topologically distinct steady mode of
propagation. 

\section{Summary and conclusion}
\label{conc}

We have considered the propagation of individual gas bubbles within a
Hele-Shaw channel containing a centred, axially-uniform constriction. We have considered the transition in behaviour as the bubble diameter is varied from the width of the rail to the width of the channel, thus bridging the two recent studies of Franco-G\'omez et al. \cite{FrancoGomez2016,FrancoGomez2017}. For
bubbles large enough that they span the entire channel width,
the behaviour of the system is the same as
the previously studied system of an air finger propagating into a
similar channel geometry \cite{Thompson2014,FrancoGomez2016}. The stable static bubble
configuration is symmetric and as the flow rate and hence capillary
number increases, the system undergoes a symmetry-breaking bifurcation
at a critical capillary number $Ca_{\rm c2}$,
above which the bubble propagates steadily in one of the deeper side channels
on either side of the constriction. The local viscous resistance 
of the channel is non-uniform and the magnitude of the induced flow
variations
increases with capillary number, which underlies the symmetry-breaking
bifurcation \cite{Thompson2014}, beyond which it is more favourable to
displace less fluid from the high-resistance, constricted
region. 

 As the bubble size decreases it does not span the entire channel
and the sidewalls can no longer stabilise the symmetric static
solution. Hence, the static bubble occupies an asymmetric position.
The introduction of axial flow can
stabilise the symmetric solution through a
symmetry-regaining pitchfork bifurcation at a critical capillary
number $Ca_{\rm c1} < Ca_{\rm c2}$. The stabilisation mechanism
involves a subtle interplay between the bubble shape, surface tension
and viscous forces and only operates when $Ca_{\rm c1} < Ca < Ca_{\rm c2}$.
We find that a depth-averaged model can accurately reproduce
the bubble shapes for asymmetric and symmetric bubbles over the range
of $Ca$ studied. The bubble speeds are less accurately captured due at least in part to the presence of films above and below the bubble in the experiment, which are neglected in the model (see \ref{A3}). As $Ca$ increases the projected areas of the
experimental bubbles becomes greater than those of the model
simulations because of the thickening of these fluid films. The percentage error in projected area increases approximately
linearly with $Ca$ and remains below 15\% for all $Ca$ studied.
As the bubble size is further reduced, the two pitchfork bifurcations move
closer together and eventually merge at a critical bubble
volume. 
Below this volume, the bubble always propagates asymmetrically
if $Ca < 0.023$, but at larger $Ca$, the bubble may break-up during its transition to an asymmetric state.

We used numerical simulations to compute both steady and unsteady
solutions and found that for higher $Ca$ there are additional
disconnected solution branches that correspond to the
Romero--Vanden-Broeck solutions, originally discovered in air-finger
flow \cite{Romero1982,VandenBroeck1983}, but
also found for finite bubbles in unbounded Hele-Shaw cells
\cite{Tanveer1987,Greenetal2017}. The results indicate that the first
of these solution branches is symmetric and initially unstable, but
restabilises at sufficiently high capillary numbers.
From the simulations, therefore,
we expect the system to be bistable in this region with the choice of
the final steady propagating state depending on the initial 
conditions. Experimentally, we find that when starting from asymmetric
initial conditions the system rapidly approaches the asymmetric
state. However, when starting from symmetric initial conditions the
system exhibits complex dynamics in which it appears to transiently
explore the weakly unstable isolated solution branches, resulting in
double-, triple- and even quadruple-tipped bubbles. These unstable
bubbles evolve ultimately leading to a topology change of the bubbles. This
scenario, in which an increasing number of weakly unstable states
develops as the flow rate increases, is reminiscent of 
the recent dynamical systems interpretation of the transition to
turbulence in shear flows
\cite{Kawahara2012,gibson_schneider_2016}. Thus it is possible that a
similar dynamic scenario occurs in the present system and provides an
underlying structure and organisation to the complex phenomena of
bubble break-up and topological change. We leave further pursuit of
these ideas for future publications.

\section*{Acknowledgements}  

A.F.-G. was funded by CONICYT. This work was supported
by the Leverhulme Trust (Grant RPG-2014-081).

\appendix

\section{Sensitivity to the rail width and sharpness}\label{A2}

   \begin{figure}[ht]
     \begin{center}
        \includegraphics[width=16.0cm]{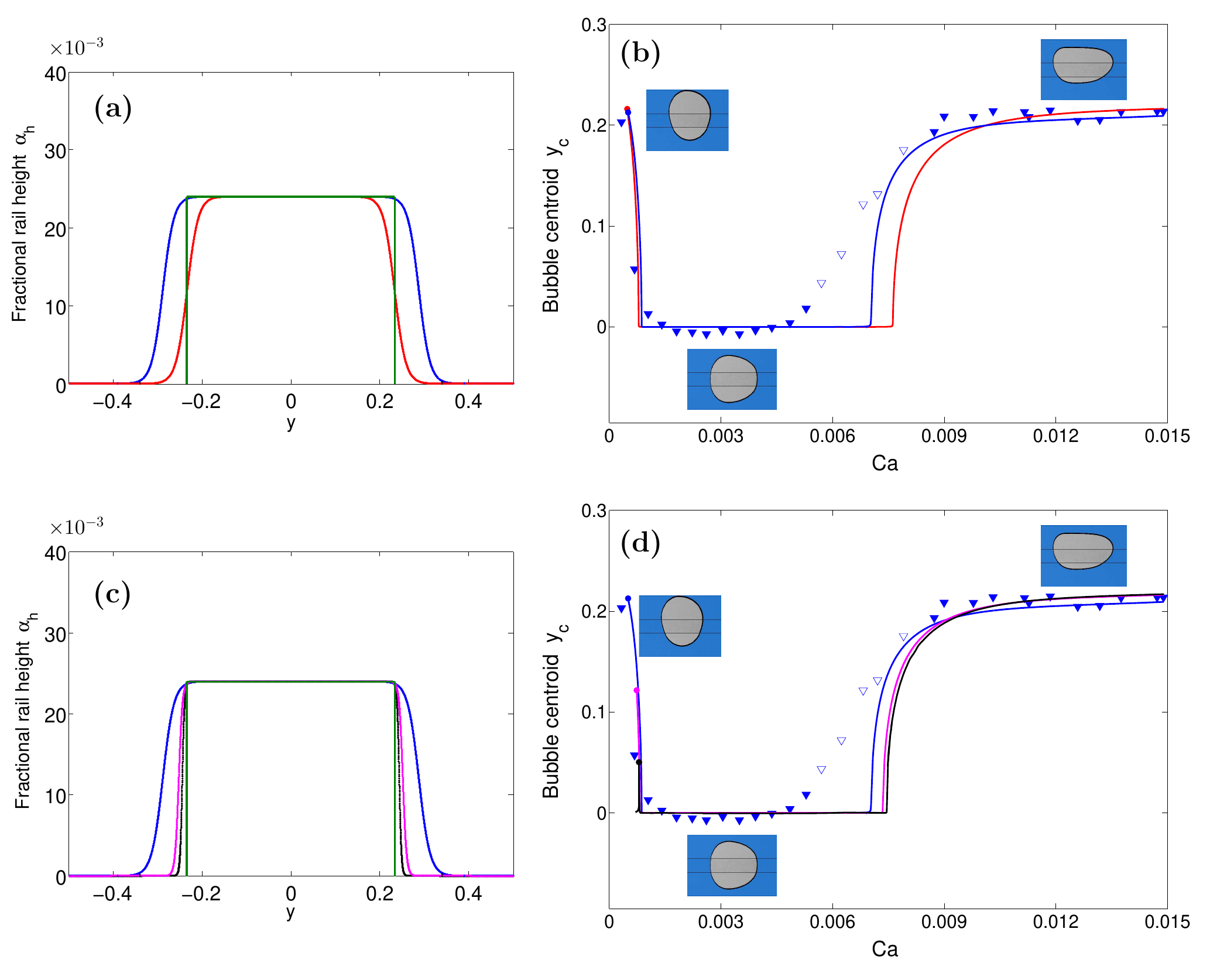}
     \end{center}
     \vspace{-0.5cm}
     \caption{\small Test of numerical rail width and sharpness for the bifurcation diagram of a propagating bubble with diameter $D=0.76$ ($D^*=22.8$ mm). The symbols (\textcolor{blue}{\large$\blacktriangledown$}) in (b) and (d) correspond to the experimental results. The measured rail profile has $\alpha_w = 0.233$, $\alpha_h = 0.024$. For the numerical model, we test various rail parameters, each with $\alpha_h = 0.024$ and $\alpha=30$: (a,b) $\alpha_w=0.288$, $s=40$ (blue), $\alpha_w=0.233$, $s=40$  (red) and (c,d) $\alpha_w=0.252$, $s=120$ (magenta) and $\alpha_w=0.244$, $s=200$  (black)}\label{fig:NumExp_rail}
   \end{figure} 

Profilometry measurements of the experimental rail reveal a geometry with approximately vertical sides, sharp edges and a rough upper surface. Averaged in the axial direction, the profile is well approximated by a rectangular shape (see supplementary material in \cite{FrancoGomez2017}), and $\alpha_w = w^*/W^*$ is measured to be $0.233$.
For finger propagation, \cite{FrancoGomez2016} found that taking $\alpha_w$ in (\ref{DepthProfile}) directly from the experimental profile gave quantitative agreement if $\alpha\geq 40$ and the sharpness parameter $s$ is fixed at $40$.
On the other hand, \cite{FrancoGomez2017} studied bubble propagation in the same channel geometry as in this paper, and found that better quantitative agreement is obtained by selecting $\alpha_w$ so that the width of the top-flat surface coincides with the width of the experimental rail; this process yields $\alpha_w = 0.288$ if $s=40$.
Figure \ref{fig:NumExp_rail}a illustrates the rail profiles, with the solid green line corresponding to the experimental rail ($\alpha_w = 6.9/30 = 0.233$, with sharp vertical sides), and two smoothed model profiles: red line ($\alpha_w = 0.233$, $s=40$) and blue line ($\alpha_w=0.288$, $s=40$).

In Figure \ref{fig:NumExp_rail}b, the experimental results for bubble
centroid as a function of capillary number for a bubble of diameter
$D=0.76$ ($D^*=22.8$ mm) are compared to the numerical results where
$\alpha_w$ in (\ref{DepthProfile}) is taken to be either $0.288$ and
$0.233$. We find that using the wider rail profile in the model
improves the agreement for the value of the second bifurcation point
and the centroid values in the asymmetric region at larger $Ca$,
though there is very little difference between the results
for the two rail widths at small $Ca$.

  The effect of sharpness is tested for $D=0.76$ using $s=40$ ($\alpha_w = 0.288$), $s=120$ ($\alpha_w=0.252$) and $s=200$ ($\alpha_w=0.244$) and the rail profiles and corresponding bifurcations are displayed in Figure \ref{fig:NumExp_rail}c, d, respectively. Each of these $s, \alpha_w$ pairs is chosen to match the width of the flat top of the rail profile.
The sharper profiles give a more realistic representation of the physical rail, but in fact the bifurcation points for the smoothest rail (blue solid line) are the closest to the experimental results. Overall, we find the bifurcations of bubble propagation do not change significantly when increasing the rail sharpness. For the calculations in the body of this paper, we take $\alpha_w = 0.288$ and $s=40$.

\section{Increase of bubble projected area with capillary number and estimate of thin film thickness}\label{A3}

\begin{figure}[ht]
    \begin{center}
     \includegraphics[width=12.5cm]{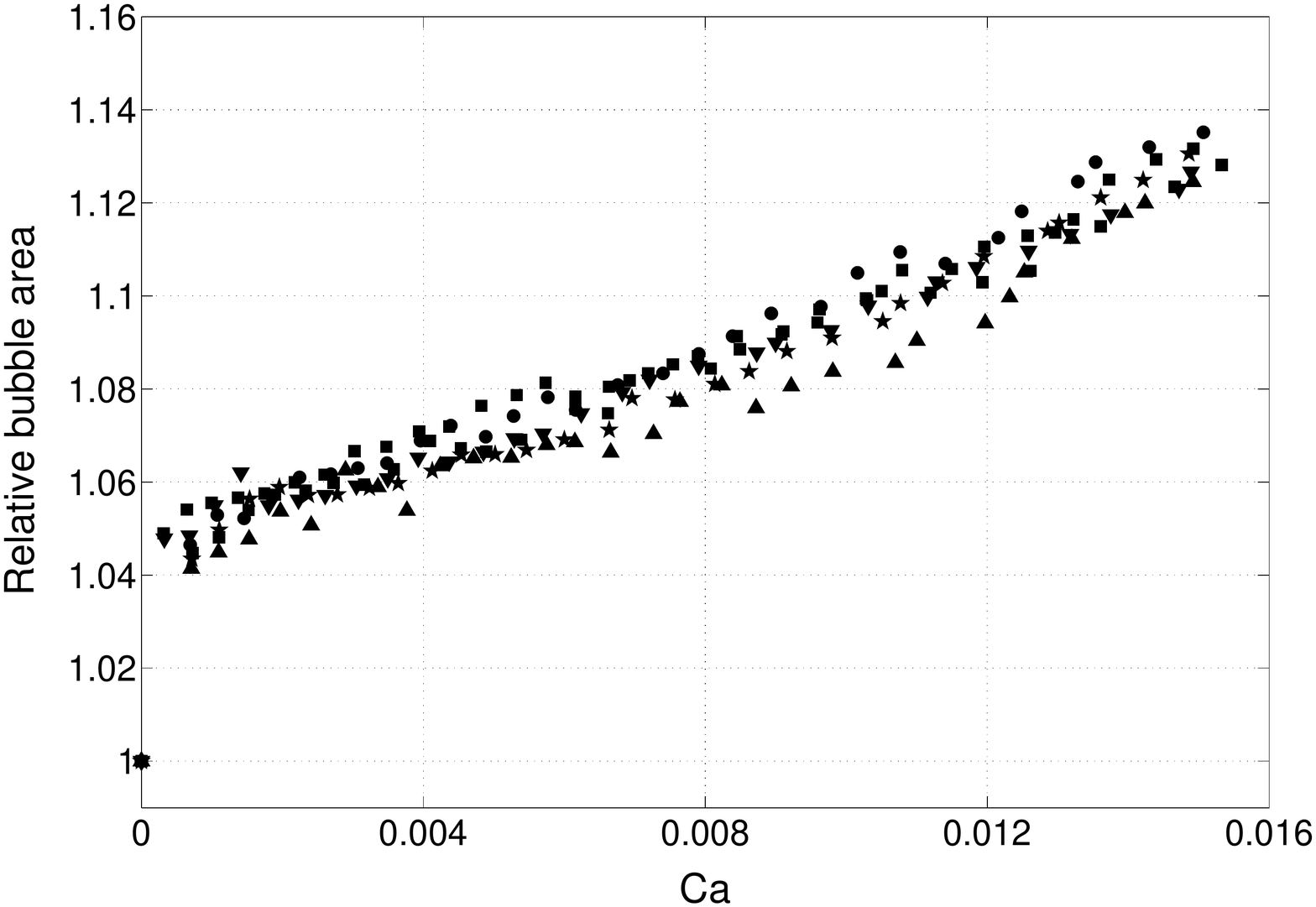}
     \end{center}
     \vspace{-0.5cm}
     \caption{\small Relative bubble area $A_{Ca}^*/A_0^*$, where $A_0^*$
       is the bubble area measured statically, as a function of the
       capillary number $Ca$; the initial condition is a centred static bubble. The plots reveal an increase of bubble area around
       $8$ \% when the flow rate is increased from $Ca=3.26\times
       10^{-4}$ ($1$ ml/min) up to $Ca=1.5\times 10^{-2}$ ($Q^*=30$
       ml/min). Note that there is a sudden increase of $5$ \% between
       the static projected area ($Ca=0$) and the area measured at
       $Ca=3.26\times 10^{-4}$ ($1$ ml/min). The change in projected
       area due to volume loss or gain along the length of the channel is better than $\pm\, 1\%$.
       The symbols indicate different bubble sizes, using the same legend as in Figure \ref{fig:ExpNumFingerOffset_lowCa}.
       }\label{fig:Area_increase_Ca}
  \end{figure}

Propagating bubbles of the diameters reported in Figure \ref{fig:ExpNumFingerOffset_lowCa} were observed to increase their projected areas due to the dependence of film thickness on capillary number. 
The relative bubble area  $A^*_{Ca}/A^*_0$, where $A^*_0$ is the bubble area measured statically, is plotted as a function of the capillary number in Figure \ref{fig:Area_increase_Ca}. 
For each bubble, there is a sudden increase of $5$ \% in the initial range of $0\leq Ca\leq 3.26\times 10^{-4}$ ($0\leq Q^*\leq 1$ ml/min).
There is followed by a slower increase of around $8$ \% over the range $3.26\times 10^{-4}\leq Ca\leq 1.5\times10^{-2}$ ($1\leq Q^*\leq 30$ ml/min).
%
%
Over the whole range of $Q^*$, the increase of area from its static value is of the order of $13$ \%. 

The normalised curves collapse inside a band of relative bubble area of approximately $\Delta (A_{Ca}^*/A_0^*)\sim 0.015$, suggesting that the increase of fluid film thickness of the reported bubbles is independent of the bubble size within the investigated range of capillary numbers. Using the data in Figure \ref{fig:Area_increase_Ca}, the average thickness of the thin films $t^*_{Ca}$ are estimated with the geometrical method previously presented by \cite{FrancoGomez2017} (supplementary material), $t^*_{Ca}=(H^*/2)\left(1-{A^*_0}/{A^*_{Ca}}\right)$. At a capillary number of $Ca=1.5\times 10^{-2}$ ($Q^*=30$ ml/min), this estimate yields fluid films of average thickness $t_{Ca}^*\sim59$ $\mu$m with a variation of $\pm\,5$ $\mu$m for the different bubble sizes.
Note that this estimate is for the film thickness averaged over the whole bubble area, and so the film may be much thinner in places.

\begin{figure}
\includegraphics[width=16cm]{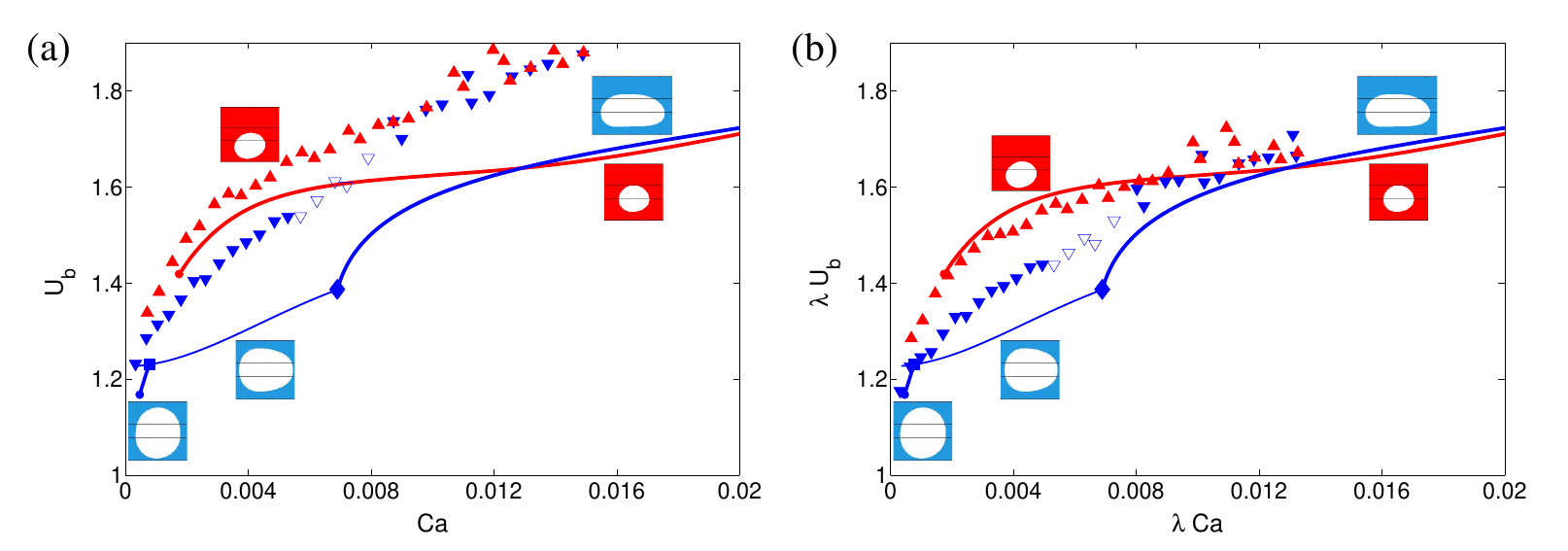}
  \caption{(a) Direct comparison of dimensionless propagation speed $U_b = Ca/Q$ as a function of $Ca$ for $D=0.50$ (red) and $D=0.76$ (blue) for original model and experiments. (b) The speed is adjusted to account for the presence of thin films by plotting $U_b \lambda $ against $Ca \lambda$, where $\lambda$ is taken to be $A^*_{0}/A^*_{Ca}$ for the experiments and $\lambda = 1$ for the model; this correction yields a closer agreement between experiment and model than the uncorrected data, except for the case of the larger bubble at small flow rates. Here the discrepancy is likely to be due to the non-uniformity of the thin films in that regime.
  \label{fig:U_Ca_comparison}}
\end{figure}

The quantitative comparisons shown earlier in this paper are for the model without any thin film corrections.  We find that predictions of the bubble shape and centroid position are in good agreement with the numerical results. However, a comparison of dimensionless bubble speed $U_b$ vs $Ca$ is in less good agreement (see Figure \ref{fig:U_Ca_comparison}a), with the experimental bubble speed exceeding that predicted by the model by up to 15\%. This discrepancy is of the same order as the increase in bubble area, and shows a broadly similar dependence on $Ca$.

As a first approximation to understand the effects of thin films, we assume that films occupy a spatially-constant fraction $\lambda$ of the channel height.
In that case, the films should appear in the kinematic boundary condition, the transverse curvature and the volume constraint. The latter two effects would in principle affect the bubble shape (which we know is in good agreement with the experiments), but the first would only lead to a rescaled $U_b$, as discussed by \cite{SaffmanTaylor1958}.
We do not have a predictive model for film thickness, so it is difficult to modify the model to match the experiments, but we can adjust the experimental results to match the dependence of the kinematic equation in the model by using the measurements of $A^*_{0}/A^*_{Ca}$ as a proxy for $\lambda$. In the model, the bubble shape is determined by $\alpha$ and $Q$ and $Ca$ is a derived parameter. We should, therefore, adjust both $U_b$ and $Ca$ to compensate for the thin films, which leads to a closer agreement between experimental and model data (see Figure \ref{fig:U_Ca_comparison}b).

\bibliographystyle{iopart-num}
\providecommand{\newblock}{}

\end{document}